\documentclass{iopart}
\def\thalf{{\textstyle{3\over 2}}}
\def\ohalf{{\textstyle{1\over 2}}}

\usepackage{epsfig}
\usepackage{graphicx}
\usepackage{hhline}
\usepackage{bm}

\begin{document}

\title{Dynamical Models of the Excitations of Nucleon Resonances}

\author{ T. Sato$^1$, T.-S. H. Lee$^2$}

\address{$^1$Department of Physics, Osaka University, Osaka, Japan}

\address{$^2$Physics Division, Argonne National Laboratory, Argonne,
 Illinois 60439, USA \\
and \\Excited Baryon Analysis Center,
Thomas Jefferson National
Accelerator Facility, Newport News, Virginia 23606, USA}
\ead{lee@phy.anl.gov,tsato@phys.sci.osaka-u.ac.jp}

\begin{abstract}
The development of a dynamical model for investigating the nucleon
resonances using the reactions of
meson production from
$\pi N$, $\gamma N$, $N(e,e')$, and $N(\nu,{\it l})$ reactions
is reviewed. The results for the $\Delta$ (1232) state
 are summarized and discussed. The progress in investigating higher mass
nucleon resonances is reported.
\end{abstract}

\section{Introduction}

The study of excited nucleon states ($N^*$) has long been recognized as
an important step towards developing a fundamental understanding of strong
interactions.  
It is an important part of the effort to understand
the structure of the nucleon since the dynamics governing the internal
structure of composite particles, such as nuclei and baryons, is closely
related to the structure of their excited states.
Within the framework of 
Quantum Chromodynamics (QCD), a clear understanding of
the spectrum and  decay scheme of the $N^*$ states will reveal
the role of  confinement and  chiral symmetry
in  the non-perturbative region.

The  $N^*$ states are unstable and
couple strongly with the meson-baryon continuum
states to form nucleon
resonances in meson production reactions on the nucleon.
Therefore the extraction of nucleon resonance parameters from the
reaction data is one of the important tasks in hadron physics.
By performing partial-wave analysis of pion-nucleon elastic scattering
data mainly during the years around 1970, many $N^*$'s have been identified.
From the resonance parameters listed by the
Particle Data Group\cite{PDG:2007} (PDG), it is clear that  
only the low-lying $N^*$ states are well 
established while there are large uncertainties in identifying higher
mass nucleon resonances. 

With the construction of high precision electron and
photon beam facilities, the situation changed 
drastically in the 1990's.
Experiments at Thomas Jefferson National 
Accelerator Facility (JLab), MIT-Bates, 
LEGS of Brookhaven National Laboratory, Mainz, Bonn, GRAAL of
Grenoble, and Spring-8 of Japan have been providing new data on the 
electromagnetic production of $\pi$, $\eta$, $K$, $\omega$, $\phi$,
and $2\pi$ final states.  These data offer a new opportunity to
to investigate $N^*$ properties, as reviewed in 
Refs.\cite{Burkert:2004sk,Lee:2006xu}.

In addition to analyzing
the world's data of meson production
from $\pi N$, $\gamma N$ and $N(e,e')$ reactions,
we  need to 
interpret the extracted 
$N^*$ parameters in terms of QCD. 
There are two possibilities.  The most 
fundamental way is to confront the extracted $N^*$ parameters 
directly with Lattice QCD calculations and QCD-based hadron structure models. 
Here the most challenging problem
is to handle the contributions from the baryon {\it continuum}  which
are coupled with the reaction channels. The second one is to develop
dynamical reaction models to analyze the meson production data. Here 
the reaction mechanisms and the internal structure of baryons
are modelled by using guidances deduced from our understanding of QCD
and many-year's study of hadron phenomenology.
In this article, we give a review of the dynamical reaction models developed in
 Refs. \cite{Sato:1996gk,Sato:2000jf,Sato:2003rq,Matsui:2005ns,
Matsuyama:2006rp,JuliaDiaz:2006xt,JuliaDiaz:2007kz,JuliaDiaz:2007fa,
Durand:2008es, Kamano:2008gr,Suzuki:2008rp}. Other approaches for 
investigating $N^*$ states  have been
reviewed in Refs.\cite{Burkert:2004sk,Lee:2006xu}.

\begin{figure}
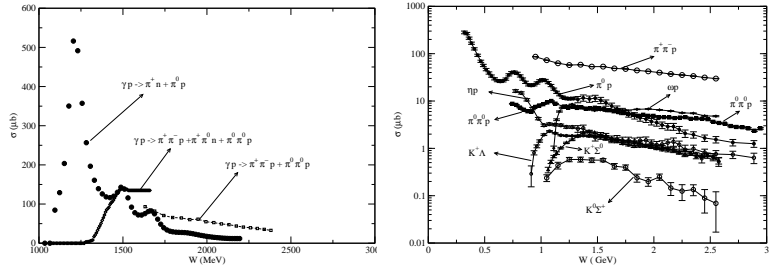

\vspace*{0.3cm}
\centering
\includegraphics[width=5cm,angle=-0]{gp_tot-1pi-2pi.eps}
\includegraphics[width=5cm,angle=-0]{ky-etan-omen-1pi-2pi.eps}
\caption{The total cross section data of meson production
in $\gamma p$ reaction.
 Left:  $1-\pi$ and
$2-\pi$ production  are compared. Right:
KY ( $K^+\Lambda$,
$K^+\Sigma^0$, $K^0\Sigma^+$), $\eta p$, and $\omega p$ production
are compared with some of the $1-\pi$ and $2-\pi$ production}
\label{fig:totcross}
\end{figure}

In practice,  the dynamical reaction
models describe the meson-baryon reaction mechanisms by
using  phenomenological Lagrangians which are constructed by using
the symmetry properties, in particular the Chiral Symmetry, deduced from
many-years' studies of meson-nucleon reactions. 
Starting from a set of phenomenological Lagrangians for mesons and baryons,
 one would ideally like to 
analyze the meson-baryon reaction data completely within the
framework of relativistic quantum field theory. The Bethe-Salpeter (BS) 
equation has been taken historically as the starting point of such
an ambitious approach.
The complications involved in solving the BS equation in the
simplest Ladder approximation have been known 
for long time.
It contains serious singularities arising from the pinching of
the integration over the time component.
In addition to the two-body unitarity cut, it has a selected
set of n-body unitarity cuts, as explained in great detail in
Refs. \cite{Afnan:1986sc,Afnan:1988fk}. Thus it is extremely difficult, if not
impossible, to apply the approach based on the Bethe-Salpeter equation
to study $N^*$ states.

Since 1990 the $\pi N$ and $\gamma N$ reactions have  been
investigated mainly by using either
the three-dimensional reductions\cite{Klein:1974aa}
of the Bethe-Salpeter equation or
the unitary transformation methods\cite{Sato:1996gk,Elmessiri:1999gy}.
These efforts were motivated mainly by the success of the meson-exchange
models of $NN$ scattering\cite{Machleidt:1989tm}, and have yielded
the meson-exchange models
developed by Pearce and Jennings\cite{Pearce:1990uj},
National Taiwan University-Argonne National Laboratory (NTU-ANL) collaboration
\cite{Lee:1991dd,Hung:2001pz}, Gross and Surya\cite{Gross:1992tj},
Sato and Lee\cite{Sato:1996gk,Sato:2000jf},
Julich Group\cite{Schutz:1994ue,Schutz:1994wp,Schutz:1998jx,Krehl:1999km},
Fuda and his collaborators\cite{Elmessiri:1999gy,Fuda:2003pd}, and
Utretch-Ohio collaboration\cite{Pascalutsa:2000bs,Caia:2005hz}.
The focus of all of these dynamical  models
was on the analysis of the data in the $\Delta$ (1232) region.
In this article, we will only review the model developed in
 Refs. \cite{Sato:1996gk,Sato:2000jf} by using the unitary transformation method.
We will also review its extension\cite{Sato:2003rq,Matsui:2005ns} to study the
$\Delta$ (1232) excitation in neutrino-induced $N(\nu,{\it l}\pi)N$
reactions. 

The main challenge of developing dynamical
reaction models of  meson production reactions
in the higher mass $N^*$ region
can be seen in Fig.\ref{fig:totcross}. We see that
two-pion photo-production cross sections shown in the left-hand-side
become larger than the one-pion photo-production
as the $\gamma p$ invariant mass exceeds $W \sim 1.4 $ GeV.
In the right-hand-side, KY ( $K^+\Lambda$,
$K^+\Sigma^0$, $K^0\Sigma^+$), $\eta p$, and $\omega p$ production
cross sections are a factor of about 10 weaker than the dominant
$\pi^+\pi^-p$ production.
From the unitarity condition, we have for any single meson production
process $\gamma N \rightarrow M B$ with $M B=\pi N, \eta N, \omega N,
  K \Lambda, K\Sigma$
\begin{eqnarray}
i(T_{MB,\gamma N} - T^*_{\gamma N,MB}) &=& \sum_{M'B'}
T^*_{ M'B',M B} \rho_{M'B'} T_{M'B',\gamma N}  \nonumber \\
& & + T^*_{\pi\pi N, M B} \rho_{\pi\pi N} T_{\pi\pi N,\gamma N}\,,
\label{eq:unitarity-0}
\end{eqnarray}
where $\rho_\alpha$ denotes an appropriate phase space factor
for the channel $\alpha$. The large two-pion production cross sections
seen in Fig.\ref{fig:totcross} indicate that the second term in the
right-hand-side of
Eq.(\ref{eq:unitarity-0}) is significant and hence the
single meson production reactions above the $\Delta$ region
must be influenced strongly by the coupling with the two-pion channels.
Similarly, the two-pion production $\gamma N \rightarrow \pi\pi N$ is
also influenced by the transition to two-body $MB$ channel
\begin{eqnarray}
i(T_{\pi\pi N,\gamma N} - T^*_{\gamma N, \pi\pi N}) &=& \sum_{M'B'}
 T^*_{ M'B',\pi\pi N} \rho_{M'B'} T_{ M'B',\gamma N}\nonumber \\
& & + T^*_{\pi\pi N, \pi\pi N} \rho_{\pi\pi N} T_{\pi\pi N,\gamma N} \,.
\label{eq:unitarity-1}
\end{eqnarray}
Clearly, a  sound dynamical reaction model must be able to
describe the two pion production and to account for the above unitarity
conditions. Such a model has been developed by using the unitary transformation
method in Ref.\cite{Matsuyama:2006rp} and
applied to investigate $\pi N$ elastic scattering\cite{JuliaDiaz:2007kz},
$\gamma N \rightarrow \pi N$ reactions\cite{JuliaDiaz:2007fa}
 $\pi N \rightarrow \eta N$ reactions\cite{Durand:2008es},
 and
$\pi N \rightarrow \pi\pi N$ reactions\cite{Kamano:2008gr}. In this article,
we will also review these results.

This article is organized as follows. In section 2, we explain
the unitary transformation method developed in Ref.\cite{Kobayashi:1997fm} using a simple
model. The constructed model Hamiltonian for investigating $N^*$ states
is given in section 3. The multi-channel multi-resonance reaction 
model developed in Refs.\cite{Sato:1996gk,Matsuyama:2006rp} for
calculating the meson-baryon
reaction amplitudes is presented
in section 4.
 In section 5, we give
formula for defining the $N$-$N^*$ transition form factors 
and calculating the cross sections of pion production
from $\pi N$, $\gamma N$, $N(e,e')$, and $N(\nu,{\it l})$
reactions. 
 The results in the
$\Delta$ (1232) region and in the higher mass $N^*$ region
are reviewed in section 6.  A summary and discussions
of future developments are given in section 7.  
 
\section{Unitary Transformation Method}

The unitary transformation method was essentially based on the same idea
of the Foldy-Wouthuysenth transformation developed in the 
study of electromagnetic
interactions. It was first developed in 1950's
by Fukuda, Sawada and Taketani \cite{fuku},
 and independently by Okubo\cite{okubo}.
This approach, called the FST-Okubo method,
has been very useful in investigating nuclear electromagnetic
currents \cite{hyuga,sko} and relativistic descriptions of
nuclear interactions \cite{glock,stno,tnso}.
The advantage of this approach is that
the resulting effective Hamiltonian
is energy independent and can readily be used in nuclear many-body calculation.

To illustrate the  unitary transformation method, 
we consider the simplest phenomenological
Lagrangian density
\begin{eqnarray}
{\it L}(x)={\it L}_0(x) + {\it L}_I(x) \,,
\label{eq:L-total}
\end{eqnarray}
where ${\it L}_0(x)$ is the usual free Lagrangians with physical masses
$m_N$ for the nucleon field $\psi_N$ and $m_\pi$ for the pion field 
$\phi_\pi$, and
\begin{eqnarray}
{\it L}_I(x) =
 \bar{\psi}_N(x)\Gamma_{N,\pi N}\psi_N(x) \phi_\pi(x) \,.
\label{eq:L-int}
\end{eqnarray}
Here $\Gamma_{N,\pi N}$ denotes 
 the  physical $\pi NN$ coupling
($ \sim f_{\pi NN}$). 
The Hamiltonian density ${\it H}(x)$ can be derived from Eqs.(\ref{eq:L-total})-(\ref{eq:L-int})by using the standard method of canonical
quantization. We then define the Hamiltonian as
\begin{eqnarray}
H = \int  {\it H}(\vec{x}, t=0)d\vec{x}.
\end{eqnarray}
The resulting Hamiltonian can be written as
\begin{eqnarray}
H=H_0 + H_I,
\label{eq:htot}
\end{eqnarray}
with 
\begin{eqnarray}
\fl
H_0 &=&\int d\vec{k}[ E_N(k) b^\dagger_{\vec{k}}b_{\vec{k}} + 
E_\pi(k)  a^\dagger_{\vec{k}}a_{\vec{k}}],
\label{eq:h0} \\
\fl
H_I&=& \Gamma_{N\leftrightarrow \pi N} \nonumber \\
\fl
   &=&\int d\vec{k}_1 d\vec{k}_2 d\vec{k} \delta (\vec{k}-
\vec{k}_1-\vec{k}_2 )[ (\Gamma_{N,\pi N}(\vec{k}_1-\vec{k}_2)
 b^\dagger_{\vec{k}}  b_{\vec{k}_1} a_{\vec{k}_2})+(h.c)],
\label{eq:hint}
\end{eqnarray}
where $b^\dagger$ and $a^\dagger$ ($b$ and $a$)
are the creation (annihilation) operators for the nucleon and
the pion, respectively. For simplicity, we drop the terms involving
the  anti-nucleon operator. 
Note that $H$ along with the other constructed generators $\vec{P}$, 
$\vec{K}$, and $\vec{J}$, as studied in Refs.\cite{glock,stno},
 define the instant-form 
relativistic quantum mechanical
description of $\pi N$ scattering.
We will work in the center of mass frame
and hence the forms of these other
generators of Lorentz group are not
relevant in the following derivations.

The essence of the unitary transformation method is to extract 
an effective Hamiltonian in a "few-body" space defined by an
unitary operator $U$, such that the resulting scattering equations can be
solved in practice.
Instead of the original equation of motion 
$H|\alpha> = E_\alpha |{\alpha}>$, we consider
\begin{eqnarray}
H^\prime |\bar{\alpha}> = E_\alpha |\bar{\alpha}>,
\end{eqnarray}
where
\begin{eqnarray}
H^\prime &=& UHU^\dagger\,, \\
|\bar{\alpha}> &=& U |\alpha>.
\end{eqnarray}
In the  approach of Kobayashi, Sato and 
Ohtsubo\cite{Kobayashi:1997fm} (KSO), 
the first step is to decompose the
interaction Hamiltonian $H_I$ Eq.(\ref{eq:hint}) into two parts
\begin{eqnarray}
H_I &= &H_I^P + H_I^Q,
\label{eq:hpq}
\end{eqnarray}
where $H_I^P$ defines the process $a \rightarrow bc$ with 
$m_a\geq m_b+m_c$ which can take place in the free space, and
$H_I^Q$ defines the virtual 
process with $m_a < m_b+m_c$.
For the simple interaction
Hamiltonian  Eq.(\ref{eq:hint}), it is clear 
that $H^P_I=0$ and $H^Q_I= H_I$.

The KSO method is to define an appropriate unitary transformation $U$
to  eliminate the virtual processes 
from transformed Hamiltonian $H^\prime$.
This can be done systematically
by using a perturbative expansion of $U$ in powers of coupling constants.
As a result the effects  of 'virtual processes' are included in the
effective operators in the transformed Hamiltonian.

Defining $U = exp(-iS)$ by a hermitian operator S and
expanding
$U = 1-iS + ...$\,, the transformed Hamiltonian can be written as
\begin{eqnarray}
\fl
H' &=& UHU^\dagger  \nonumber \\
\fl
&=& U(H_0 + H^P_I + H^Q_I ) U^\dagger  \nonumber \\
\fl
&=& H_0 + H^{P}_I + H^{Q}_I + [H_0 , iS\,] 
+ [H_I , iS\,] + {1\over 2!} \, \Big[ [H_0 , iS\,] , iS\,\Big] + \cdots .
\label{eq:hpq1}
\end{eqnarray}
To eliminate from Eq.(\ref{eq:hpq1}) the virtual processes which are of first-order in 
the coupling constant,  the KSO method imposes the condition
that

\begin{equation}
H^{Q}_{I} + [H_0 , iS\,] = 0 \,.
\label{eq:hpq-s}
\end{equation}
Since $H_0$ is a diagonal operator in Fock-space , 
Eq.(\ref{eq:hpq-s}) clearly implies that $iS$ must have the same
operator structure of $H^{Q}_{I}$ and is first order in coupling constant.  
By using Eq.(\ref{eq:hpq-s}), Eq.(\ref{eq:hpq1}) can be written as
\begin{eqnarray}
H^\prime = H_0 + H_I^\prime \,, 
\end{eqnarray}
with
\begin{eqnarray}
H_I^\prime = H_I^P + [H_I^P,iS\,] + \frac{1}{2} \,[H_I^Q,iS\,]
+ \mbox{higher order terms}\,.
\label{eq:hpq2}
\end{eqnarray}
Since $H_I^P$, $H_I^Q$, and $S$ are 
all of the first order in the coupling constant, 
all processes included in the second and third terms of
the $H_I^\prime$ are of the second order in coupling constants.

We now turn to illustrating how the constructed $H_I^\prime$ of
Eq.(\ref{eq:hpq2}) can be used to describe the $\pi N$ scattering
if the higher order terms are dropped. We consider
the simple Hamiltonian defined by Eqs.(\ref{eq:htot})-(\ref{eq:hint}) which
gives $H^P_I=0$ and $H^Q_I=\Gamma_{N\leftrightarrow \pi N}$.
Our first task is to find $S$ by solving Eq.(\ref{eq:hpq-s}) within the
Fock space spanned by the eigenstates of $H_0$
\begin{eqnarray}
& &H_0|N> = m_N|N>\,, \\
& &H_0|\vec{k},\vec{p} > = (E_\pi(k)+E_N(p))|\vec{k},\vec{p}>\,, \\
& &H_0 | \vec{k}_1,\vec{k}_2,\vec{p} > = 
((E_\pi(k_1)+E_\pi(k_2)+ E_N(p)) | \vec{k}_1,\vec{k}_2,\vec{p} >\,, \\
& & \cdot\cdot\cdot\cdot \nonumber 
\end{eqnarray}
For two eigenstates $f$ and $i$ of $H_0$, the solution
of Eq.(\ref{eq:hpq-s}) clearly is
\begin{eqnarray}
<f|(iS)|i> = -\frac{<f|H^Q_I|i>}{E_f- E_i} \,.
\end{eqnarray}
For the considered $H^Q_I=\Gamma_{N\leftrightarrow \pi N}$
we thus get the following non-vanishing matrix elements
\begin{eqnarray}
<\vec{k}, \vec{p} |(iS)|N> &=& - \Gamma_{N,\pi N}(k)
\frac{\delta(\vec{k}+\vec{p})}{E_\pi(k)+E_N(p)-m_N}\,,
\label{eq:r-1} \\
<N|(iS)|\vec{k}^{\,\prime},\vec{p}^{\,\prime}> &=&
- \frac{\delta(\vec{k}^{\,\prime}+\vec{p}^{\,\prime})}{m_N-E_\pi(k')
-E_N(p')}\Gamma^*_{N,\pi N}(\vec{k}^{\,\prime}) \,,
\label{eq:r-2}  
\end{eqnarray}
and
\begin{eqnarray}
\fl
<\vec{k}_1,\vec{k}_2,\vec{p}|(iS)|\vec{k}^\prime ,\vec{p}^{\,\prime}> 
&=&
\frac{-\delta(\vec{k}'-\vec{k}_2)
\delta(\vec{p}'-\vec{k}_1-\vec{p}) \Gamma^*_{N,\pi N}(k_1)}
{E_\pi(k_1)+E_\pi(k_2)+E_N(p)-E_\pi(k')
-E_N(p')} + (1 \leftrightarrow 2) \label{eq:r-3}\nonumber \\
\fl
&=&\Gamma^*_{N,\pi N}(k_1)
\frac{-\delta(\vec{k}'-\vec{k}_2)\delta(\vec{p}'-\vec{k}_1-\vec{p})}
{E_\pi(k_1)+E_N(p)-E_N(p')}  + (1 \leftrightarrow 2)\,,
\label{eq:r-4} \\
& & \nonumber \\
\fl
<\vec{k},\vec{p}|(iS)|\vec{k}_1,\vec{k}_2,\vec{p}> 
&=&
\frac{-\delta(\vec{k}-\vec{k}_1)\delta(\vec{p}'-\vec{k}_2-\vec{p})
\Gamma_{N,\pi N}(k_2)
}
{E_\pi(k)+E_N(p)-E_\pi(k_1)-E_\pi(k_2)
-E_N(p)} + (1 \leftrightarrow 2) \nonumber \\
\fl
&=&\Gamma_{N,\pi N}(k_2)
\frac{-\delta(\vec{k}-\vec{k}_1)\delta(\vec{p}'-\vec{k}_2-\vec{p})}
{E_N(p)-E_\pi(k_2)-E_N(p)}
 + (1 \leftrightarrow 2)
\,. \label{eq:r-5}
\end{eqnarray}
With the above matrix elements and recalling that $H^P_I=0$ and 
$H^Q_I=\Gamma_{N\leftrightarrow \pi N}$ for
the considered simple case, the matrix element of the effective Hamiltonian
Eq.(\ref{eq:hpq2}) in the center of mass frame ($\vec{p}=-\vec{k}$
and $\vec{p}'=-\vec{k}'$) is
\begin{eqnarray}
<\vec{k}| H_I^\prime |\vec{k}'> &=&
\frac{1}{2}\sum_{I}[(<\vec{k}|\Gamma_{N\leftrightarrow \pi N}|I><I|
(iS)|\vec{k}'> \nonumber \\
& & - <\vec{k}|(iS)|I><I|\Gamma_{N\leftrightarrow \pi N}|\vec{k}'>]
\,.
\end{eqnarray}
The only possible intermediate states are 
$|I> = |N> + |\pi(k_1)\pi(k_2) N(P_I)>$. By using
 Eqs.(\ref{eq:r-1})-(\ref{eq:r-5}) we then obtain
\begin{eqnarray}
<\vec{k}| H_I^\prime |\vec{k}'> &=&
 v^{(s)}(\vec{k},\vec{k}')+v^{(u)}(\vec{k},\vec{k}') \,.
\label{eq:veff}
\end{eqnarray}
where
\begin{eqnarray}
v^{(s)}(\vec{k}, \vec{k}')&=&\frac{1}{2}\Gamma^*_{N,\pi N}(k)
[\frac{1}{E_\pi(k)+E_N(k)-m_N}\nonumber \\
& & + \frac{1}{E_\pi(k')+E_N(k')-m_N}]
\Gamma^*_{N,\pi N}(k')\,,
\label{eq:veff-s} \\
v^{(u)}(\vec{k}, \vec{k}')&=&\frac{1}{2}\Gamma^*_{N,\pi N}(k')
[\frac{1}{E_N(k)-E_\pi(k')-E_N(\vec{k}+\vec{k}')} \nonumber \\
& &+ \frac{1}{E_N(k')-E_\pi(k)-E_N(\vec{k}+\vec{k}')}]\Gamma_{N,\pi
 N}(k) \,.
\label{eq:veff-u} 
\end{eqnarray}

Note that up to the same order Eq.(\ref{eq:veff}) should have
an additional term which is the one-pion-loop contribution to
the single nucleon state. 
Such a mass renormalization term is dropped in practice,
since it is part of the physical nucleon mass in the resulting effective 
Hamiltonian.
If we treat this mass renormalization explicitly, we then will not get
a solvable few-body problem, but a many-body problem which is as complicated as
the original field theory problem.
We also note that
$v^{(s)}$ of Eq.(\ref{eq:veff-s}) is due to the intermediate
"physical" nucleon state state $|I> = |N>$.
This is the consequence of the unitary
transformation which eliminates the
"virtual" $\pi N \leftrightarrow N$ process.
Here we see an important difference between 
 $v^{(s)}$  and the so-called nucleon-pole term from approaches based on some
models based on the three-dimensional reduction of Bethe-Salpeter equations
 and the time-order perturbation theory\cite{Krehl:1999km}.
There is no bare mass $m^0_N$
and energy-dependence in $v^{(s)}$.

With the above derivations,
the effective Hamiltonian Eq.(\ref{eq:hpq2}) can be explicitly written as
\begin{eqnarray}
H'= H_0 + V\,,
\label{eq:effh-0}
\end{eqnarray}
where
\begin{eqnarray}
H_0 &=&\int d\vec{k}[ E_N(k) b^\dagger_{\vec{k}}b_{\vec{k}} +
E_\pi(k)  a^\dagger_{\vec{k}}a_{\vec{k}}]\,,
\label{eq:eff-h0} \\
V &=& \int d\vec{k}d\vec{k}' [v^{(s)}(\vec{k}, \vec{k}')+v^{(u)}(\vec{k}, \vec{k}')]
a^\dagger_{\vec{k}}b^\dagger_{-\vec{k}}
a_{\vec{k}'}
b_{-\vec{k}'} \,. \label{eq:veff-0}
\end{eqnarray}

To see the analytic properties of the reaction amplitudes based on the
effective Hamiltonian  Eq.(\ref{eq:effh-0}),
let us first recall
how the bound states and resonances are defined in a Hamiltonian formulation.
In operator form the reaction amplitude
is defined by
\begin{eqnarray}
t(E) = V + V\frac{1}{E-H_0+i\epsilon}t(E)\,, \label{eq:lseq}
\end{eqnarray}
or
\begin{eqnarray}
t(E) = V + V\frac{1}{E-H'+i\epsilon}V\,. \label{eq:loweq0}
\end{eqnarray}
The analytic structure of scattering amplitude can be most transparently
seen by using the spectral expansion of the Low equation Eq.(\ref{eq:loweq0})
\begin{eqnarray}
<k'|t(E)|k> &=&
 <k'|V|k>+ \sum_{i} \frac{<k'|V|\Phi_{\epsilon_i}><\Phi_{\epsilon_i}|V|k>}
{E - \epsilon_i} \nonumber \\
& & +
\int_{E_{th}}^{\infty} \frac{<k'|V|\Psi^{(+)}_{E'}><\tilde{\Psi}^{(+)}_{E'}|V|k>}
{E-E'+i\epsilon} \,, \nonumber \\
\label{eq:c1}
\end{eqnarray}
where  $E_{th}$ is the threshold of the
reaction channels,  $\Phi_{\epsilon_i}$ and $\Psi^{(+)}_{E'}$ are the discrete
bound states and the scattering states, respectively.
They form a complete set and satisfy
\begin{eqnarray}
H'|\Phi_{\epsilon_i}> &=& \epsilon |\Phi_{\epsilon_i}>\,, \label{eq:heff-bs} \\
H'|\Psi^{(+)}_{E'}> &=& E' |\Psi^{(+)}_{E'}>\,. \label{eq:heff-scatt}
\end{eqnarray}
Of course bound state energies $\epsilon_i$
 are below the production threshold $E_{th}$. We now note that
because of the two-body nature of $V$  defined by Eq (\ref{eq:veff-0}),
 Eq.(\ref{eq:heff-bs}) has the one-nucleon solution
$H'|N> = H_0|N> = m_N |N>$. But it does not contribute to the
second term of Eq.(\ref{eq:c1}) because $<\pi N|V|N> =0$.
Thus the amplitude Eq.(\ref{eq:c1}) 
 does not have a nucleon pole which corresponds to
bound state with a mass of physical nucleon and is formed
by the $physical$ $N$ and $\pi$ of the starting Lagrangian
Eq. (\ref{eq:L-total}).
This is consistent with the experiment. 
Clearly, our approach is very different from the S-matrix approach which
requires that the $\pi N$ scattering amplitude must have a pole at
$E=m_N$.
Similar feature is  also obtained by using
the unitary transformation of Shebeko et al.\cite{Shebeko:2001bf,Korda:2004sy}.

To end this section, we mention that the unitarity condition
only requires that an acceptable model must have unitarity cut in
physical region $E \geq m_\pi + m_N$.
 This is trivially satisfied  in the 
the model defined by the effective Hamiltonian 
Eqs.(\ref{eq:eff-h0})-(\ref{eq:veff-0}) since the interaction $V$ is energy
independent.
This is an important advantage in applying the method of unitary transformation
to develop a multi-channels multi-resonances reaction models for
investigating meson-nucleon reactions in the nucleon resonance region, 
as developed in Ref.\cite{Matsuyama:2006rp}.
 In a model with an energy-dependent $V$ such as the Julich model\cite{Krehl:1999km} 
the unitarity condition is much more difficult to satisfy, and
the analytic continuation of the scattering t-matrix defined by 
Eqs.(\ref{eq:c1}) to complex $E$-plane is in general much more complex.

\section{Model Hamiltonian}

With the unitary transformation method explained in section 2, it is
straightforward to derive 
 a  model Hamiltonian for
constructing a coupled-channel reaction model with 
$\gamma N$, $\pi N$, $\eta N$ and $\pi\pi N$ channels.
Since significant parts of the $\pi\pi N$ production are known
experimentally to be through 
the unstable states $\pi\Delta$, $\rho N$, and
perhaps also $\sigma N$, we will also include  $bare$ $\Delta$,
$\rho$ and $\sigma$ degrees of freedom in our formulation. Furthermore,
we introduce $bare$ $N^*$ states to represent 
the quark-core components of the nucleon resonances.
The model is expected to be valid up to $W=2$ GeV below which three
pion production is very weak.

The starting point is a set of Lagrangians 
describing the interactions between mesons 
($M =\gamma, \pi,\eta ,\rho, \omega, \sigma \cdot\cdot\cdot$) and 
baryons ($B = N, \Delta, N^* \cdot\cdot\cdot$). These Lagrangian
are constrained by various well-established symmetry properties,
such as the invariance under isospin, parity, and gauge
transformation. The chiral symmetry is also implemented as 
much as we can.
The considered Lagrangians are given in Ref.\cite{Matsuyama:2006rp}. 
For completeness,  we recall  in Appendix A 
parts of these Lagrangians which were used
in investigating the $\Delta$ (1232) resonance.

By applying the
standard canonical quantization, we obtain 
a  Hamiltonian of the following form 
\begin{eqnarray}
H &=&\int  {\it h(\vec{x},t=0)} d\vec{x} \nonumber \\
&=& H_0 + H_I\,,
\label{eq:sl-1}
\end{eqnarray}
where ${\it h(\vec{x},t)}$ is the Hamiltonian density constructed from
the starting Lagrangians and the conjugate momentum field operators.
In Eq.(\ref{eq:sl-1}), $H_0$ is the free Hamiltonian and
\begin{eqnarray}
H_I = \sum_{M,B,B^\prime} \Gamma_{MB\leftrightarrow B^\prime}
+\sum_{M,M',M''} h_{M'M''\leftrightarrow M} \,,
\label{eq:sl-2}
\end{eqnarray}
where 
 $ \Gamma_{MB \leftrightarrow B^\prime}$ describes
the absorption and emission of a meson($M$) by a baryon($B$) such
as $\pi N \leftrightarrow N$ and $\pi N \leftrightarrow \Delta$, and
$h_{M'M''\leftrightarrow M} $ describes the vertex interactions between
mesons such
as $\pi\pi \leftrightarrow \rho$ and $\gamma \pi \leftrightarrow \pi$.

\begin{figure}[t]
\centering
\includegraphics[width=8cm,angle=-0]{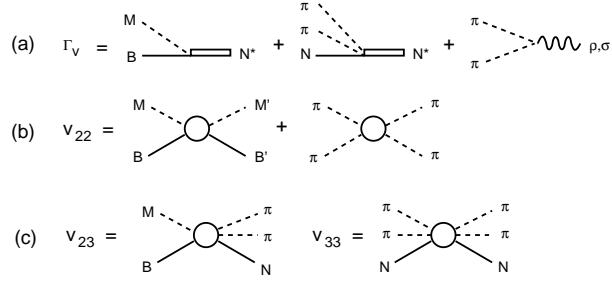}
\caption{Basic mechanisms of the Model Hamiltonian defined in Eqs.(42)-(44). }
\label{fig:h}
\end{figure}

Our main step is to derive from 
Eqs.(\ref{eq:sl-1})-(\ref{eq:sl-2}) an effective Hamiltonian
which contains interactions involving $\pi\pi N$ three-particle
states.  This is accomplished by
applying the unitary transformation 
method up to the
third order in interaction $H_I$ of Eq.(\ref{eq:sl-2}). The resulting
effective
Hamiltonian is of the following form
\begin{eqnarray}
H_{eff}= H_0 + V \,,
\label{eq:H-1}
\end{eqnarray}
with 
\begin{eqnarray}
H_0 = \sum_{\alpha} K_\alpha \,,
\end{eqnarray}
where $K_\alpha = \sqrt{m_\alpha^2+\vec{p_\alpha}^2}$
is the free energy operator of particle $\alpha$ with a mass $m_\alpha$,
and the interaction Hamiltonian is 
\begin{eqnarray}
V =\Gamma_V + v_{22}   + v' \,,
\label{eq:H-2}
\end{eqnarray} 
where 
\begin{eqnarray}
\Gamma_V&=& \{ \sum_{N^*}
(\sum_{MB} \Gamma_{N^*\rightarrow MB} +\Gamma_{N^* \rightarrow \pi\pi N})
+ \sum_{M^*} h_{M^*\rightarrow \pi\pi} \} + \{h.c.\} \,, 
\label{eq:gammav} \\ 
v_{22}&=& \sum_{MB,M^\prime B^\prime}v_{M^\prime B^\prime,MB}
+v_{\pi\pi} \,.
\label{eq:v22} 
\end{eqnarray}
Here ${h.c.}$ denotes the hermite conjugate of the terms on its 
left-hand-side.
In the above equations, 
 $MB =  \gamma N, \pi N, \eta N, \pi\Delta, 
\rho N, \sigma N$ represent the considered meson-baryon states.
The resonance associated with
the $bare$ baryon state $N^*$ is
induced by the vertex interactions 
$\Gamma_{N^* \rightarrow MB }$ and $\Gamma_{N^* \rightarrow \pi\pi N}$. 
Similarly, the $bare$
meson states $M^* $ = $\rho$, $\sigma$ can develop into resonances through
the vertex interaction $h _{M^*\rightarrow \pi\pi}$. These vertex
interactions  are illustrated in
Fig.\ref{fig:h}(a).
Note that the masses $M^0_{N^*}$ and $m^0_{M^*}$ 
of the bare  states $N^*$ and $M^*$ 
are the parameters of the model which will be determined by fitting the
$\pi N$ and $\pi\pi$ scattering data. They differ
from the empirically determined
resonance positions by mass shifts which are due to the
coupling of the bare states with the meson-baryon
$scattering$ states. It is thus reasonable to speculate that
these bare masses can be
identified with the mass spectrum predicted by  the hadron structure 
calculations which do not account for the meson-baryon $continuum$ 
scattering states, such as the calculations
based on the constituent quark models which do not have
meson-exchange quark-quark interactions. It is however much more difficult,
but more interesting, to relate these bare masses to the $current$
Lattice QCD calculations which
can not account for the scattering states rigorously mainly
because of the limitation of the lattice spacing.

\begin{figure}[h]
\centering
\includegraphics[width=8cm]{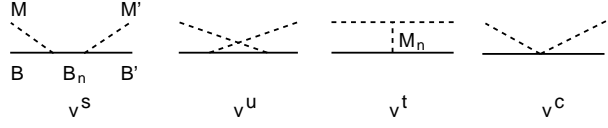}
\caption{Mechanisms for $v_{M'B',MB}$ of Eq. (\ref{eq:v22}):
(a) direct s-channel, (b) crossed u-channel,
 (c) one-particle-exchange t-channel, (d) contact interactions.}
\label{fig:mbmb}
\end{figure}

\begin{figure}[b]
\centering
\includegraphics[width=12cm,angle=-0]{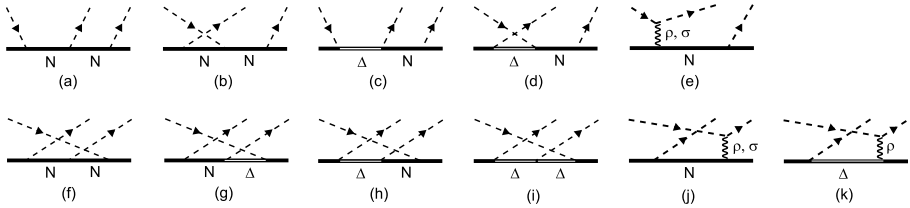}
\caption{The considered $v_{\pi N,\pi\pi N}$ of $v_{23}$ of Eq.(44). }
\label{fig:mbpipin}
\end{figure}

In Eq.(\ref{eq:v22}), $v_{M^\prime B^\prime,MB}$ is the
non-resonant meson-baryon interaction and $v_{\pi\pi}$ is the
non-resonant $\pi\pi$ interaction. They are illustrated in Fig.\ref{fig:h}(b).
The third term in Eq.(\ref{eq:H-2}) describes the non-resonant
interactions involving $\pi\pi N$ states
\begin{eqnarray}
v' =  v_{23} + v_{33}  \,,
\label{eq:v23}
\end{eqnarray}
with 
\begin{eqnarray}
v_{23}&=&\sum_{MB}[( v_{\pi\pi N,MB}) + (h.c.)]\,, \nonumber \\
v_{33}&=&  v_{\pi\pi N, \pi\pi N}\,. \nonumber
\end{eqnarray}
They are illustrated in Fig.\ref{fig:h}(c).
All of these interactions are defined by the tree-diagrams generated from
the considered Lagrangians. They are illustrated in Fig.\ref{fig:mbmb}
for two-body interactions $v_{M'B',MB}$ and in Fig.\ref{fig:mbpipin}
for $v_{\pi\pi N,MB}$. In practice, we neglect $v_{\pi\pi}$
 and $v_{\pi\pi N ,\pi\pi N}$. We also only consider $v_{\pi\pi N,\pi N}$
and $v_{\pi\pi N,\gamma N}$ of $v_{\pi\pi N,MB}$.
These two interactions are illustrated in
Fig.\ref{fig:mbpipin}.
The calculations of the
matrix elements of these interactions were explained in 
details in Ref. \cite{Matsuyama:2006rp}.
Here we only mention that the matrix elements of
these interactions are calculated from the usual
Feynman amplitudes with 
the energies of off-mass-shell particles in the intermediate states 
defined by the three momenta
of the initial and
final states, as specified by the unitary transformation methods.
Thus they are independent of the collision energy $E$.

\section{Multi-channels Multi-resonances Reaction Model}

Our next task is to derive a set of dynamical coupled-channel equations
 for describing $\gamma N, \pi N \rightarrow MB$ reactions
within the model space $N^*\oplus MB \oplus \pi\pi N$.
The starting point is the Lippman-Schwinger equation for  the scattering T-matrix
\begin{eqnarray}
<a|T(E)|b>=<a|V|b> + <a|V\frac{1}{E-H_0 + i\epsilon} T(E)|b>  \,,
\label{eq:loweq}
\end{eqnarray}
where the interaction $V$ is defined from the effective Hamiltonian
in Eqs.(\ref{eq:H-1})-(\ref{eq:v23}).
We choose the normalization that the T-matrix is related to the S-matrix by
\begin{eqnarray}
<a|S(E)|b> = \delta_{ab}- 2\pi i \delta^4(P_a - P_b)<a|T(E)|b> \,.
\label{eq:smatrix}
\end{eqnarray}
Since the interaction $V$, defined by Eqs.(\ref{eq:H-2})-(\ref{eq:v23}),
is energy independent, it is rather straightforward to follow
the formal scattering theory given in Ref.\cite{Goldberger:1964} 
to show that Eq.(\ref{eq:loweq}) leads to the following
unitarity condition 
\begin{eqnarray}
\fl
<a|T(E)-T^\dagger(E)|b> = -2\pi i\sum_{c}
<a|T^\dagger(E)|c>\delta(E_c-E)<c|T(E)|b> \,,
\label{eq:unitarity}
\end{eqnarray}
where $a,b,c$ are the
reaction channels in the considered energy region. 

We cast Eq. (\ref{eq:loweq}) into a more convenient form
for practical calculations. In the derivations, the unitarity condition
Eq.(\ref{eq:unitarity}) must be maintained exactly.
We achieve this rather complex task by
applying the standard projection operator techniques\cite{Feshbach:1992},
 similar to that employed in 
a study of $\pi NN$ scattering\cite{Lee:1985jq}.
The details of our derivations 
are given in Appendix B of Ref. \cite{Matsuyama:2006rp}.
To explain our coupled-channel equations, it is
sufficient to present the formula obtained from setting 
$\Gamma_{N^*\rightarrow \pi\pi N} =0 $ in our derivations.
Here we explain these equations and discuss their dynamical content. 

\begin{figure}
\centering
\includegraphics[width=10cm,angle=-0]{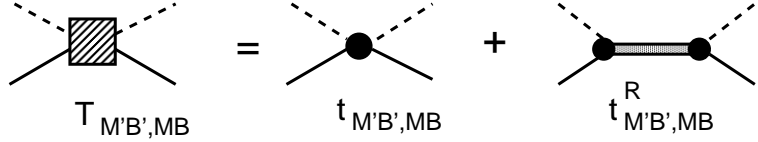}
\caption{ Graphical representations of Eq. (\ref{eq:tmbmb}).}
\label{fig:tmatmbmb-1}
\end{figure}

\begin{figure}
\centering
\includegraphics[width=12cm,angle=-0]{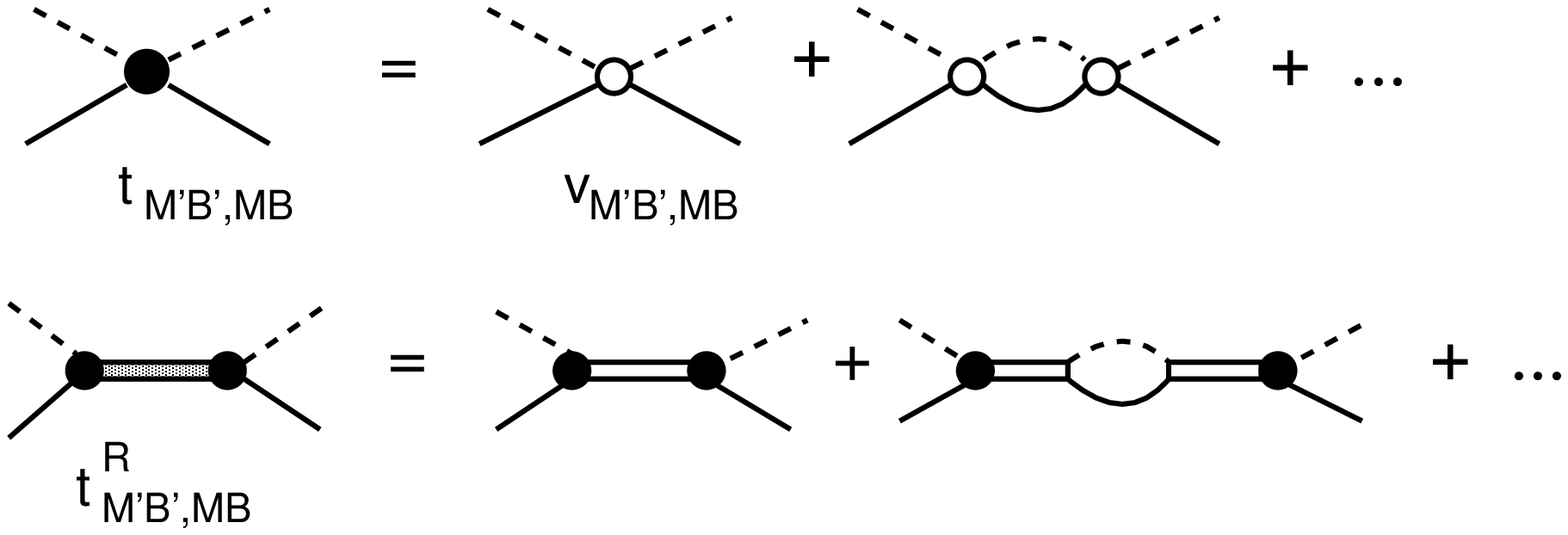}
\caption{ Graphical representations of Eqs. (\ref{eq:tmbmb-r}) and (\ref{eq:cc-mbmb}).}
\label{fig:tmatmbmb-2}
\end{figure}

The  resulting $MB \rightarrow M^\prime B^\prime$   amplitude 
$T_{M^\prime B^\prime,MB}$  
in each partial wave 
consists of a non-resonant amplitude $t_{M^\prime B^\prime,MB}(E)$
and a resonant amplitude $t^R_{M^\prime B^\prime,MB}(E)$
as illustrated in Figs. \ref{fig:tmatmbmb-1} and \ref{fig:tmatmbmb-2}. It
can be written as
\begin{eqnarray}
T_{M^\prime B^\prime,MB}(E)  &=&  t_{M^\prime B^\prime,MB}(E)
+ t^R_{M^\prime B^\prime,MB}(E) \,.
\label{eq:tmbmb}
\end{eqnarray}

\begin{figure}
\centering
\includegraphics[width=8cm,angle=-0]{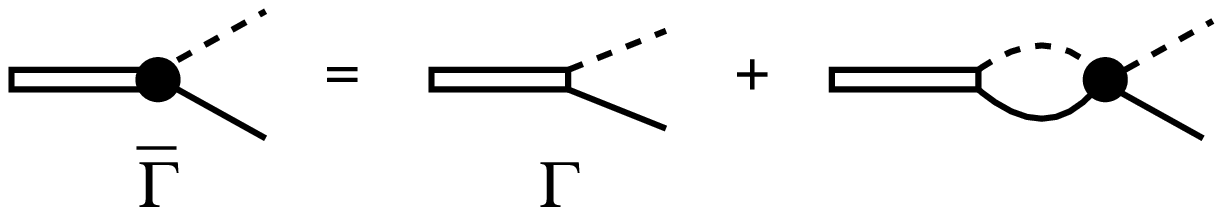}
\caption{ Graphical representations of Eqs. (\ref{eq:mb-nstar})-(\ref{eq:nstar-mb}).}
\label{fig:g-dress}
\end{figure}

The second resonant term in the right-hand-side of
Eq.(\ref{eq:tmbmb}) is defined by
\begin{eqnarray} 
t^R_{M^\prime B^\prime,MB}(E)= \sum_{N^*_i, N^*_j}
\bar{\Gamma}_{N^*_i \rightarrow M^\prime B^\prime}(E)
 [D(E)]_{i,j}
\bar{\Gamma}_{MB \rightarrow N^*_j}(E)
 \,,
\label{eq:tmbmb-r} 
\end{eqnarray}
with
\begin{eqnarray}
[D(E)^{-1}]_{i,j}(E) = (E - M^0_{N^*_i})\delta_{i,j} - \bar{\Sigma}_{i,j}(E)\,,
\label{eq:nstar-g}
\end{eqnarray}
where $M_{N^*}^0$ is the  mass of a bare $N^*$ state, and
the self-energies are 
\begin{eqnarray}
\bar{\Sigma}_{i,j}(E)= \sum_{MB}
\bar{\Gamma}_{MB \rightarrow N^*_i}(E)
 G_{MB}(E)
\Gamma_{N^*_j\rightarrow MB}
 \,.
\label{eq:nstar-sigma}
\end{eqnarray}
In general, the bare states mix with each other
through the off-diagonal matrix elements of the self-energies.
The dressed vertex interactions in Eq. (\ref{eq:tmbmb-r}) and
Eq. (\ref{eq:nstar-sigma}) illustrated in Fig. \ref{fig:g-dress} are 
 (defining
 $\Gamma_{MB\rightarrow N^*}=\Gamma^\dagger_{N^* \rightarrow MB}$)
\begin{eqnarray}
\fl
\bar{\Gamma}_{MB \rightarrow N^*}(E)  =  
  { \Gamma_{MB \rightarrow N^*}} + \sum_{M^\prime B^\prime}
\Gamma_{M^\prime B^\prime \rightarrow N^*}
G_{M^\prime B^\prime}(E)
t_{M^\prime B^\prime,MB}(E)
\,, 
\label{eq:mb-nstar}
\\
\fl
\bar{\Gamma}_{N^* \rightarrow MB}(E)
 =  \Gamma_{N^* \rightarrow MB} +
\sum_{M^\prime B^\prime} 
t_{M B,M^\prime B^\prime}(E) 
G_{M^\prime B^\prime }(E)
\Gamma_{N^*\rightarrow M^\prime B^\prime}
\,. 
\label{eq:nstar-mb}
\end{eqnarray}
The meson-baryon propagator $G_{MB}$ in the above equations 
takes the following form
\begin{eqnarray}
G_{MB}(E)=\frac{1}{E - K_B -K_M  -\Sigma_{MB}(E) + i\epsilon} \,,
\label{eq:g-mb-1}
\end{eqnarray}
where the  mass shift $\Sigma_{MB}(E)$ depends on the considered $MB$
channel.
It is $\Sigma_{MB}(E)=0$ for the
stable particle channels  $MB=\pi N, \eta N$.
For channels containing an
unstable particle,  such as $MB=\pi\Delta, \rho N, \sigma N$, we have
\begin{eqnarray}
\fl
\Sigma_{MB}(E) =[ <MB| g_V
\frac{P_{\pi\pi N}}{E-K_\pi-K_\pi -K_N+i \epsilon}
g^\dagger_V |MB> ]_{un-connected}\,, 
\label{eq:sigma-m}
\end{eqnarray}
with
\begin{eqnarray}
g_V =\Gamma_{ \Delta\rightarrow \pi N}+
h_{\rho\rightarrow \pi\pi} + h_{\sigma \rightarrow \pi\pi} \,.
\label{eq:gv}
\end{eqnarray}
In Eq.(\ref{eq:sigma-m}) "$un-connected$" means that the
stable particle, $\pi$ or $N$, of the $MB$ state is a spectator in
 the $\pi\pi N$ propagation. Thus $\Sigma_{MB}(E)$ is just the mass
renormalization of the unstable particle in the $MB$ state.
It is important to note that the
resonant amplitude $t^R_{M'B',MB}(E)$ is influenced by the non-resonant
amplitude $t_{M'B',MB}(E)$, as seen
in Eqs. (\ref{eq:tmbmb-r})-(\ref{eq:nstar-mb}).

The non-resonant amplitudes $ t_{M^\prime B^\prime,MB}$ 
in Eq.(\ref{eq:tmbmb}) and Eqs.(\ref{eq:mb-nstar})-(\ref{eq:nstar-mb})
are defined by the following coupled-channel equations
\begin{eqnarray}
\fl
t_{M^\prime,B^\prime,MB}(E)= V_{M^\prime B^\prime,MB}(E) 
+ \sum_{M^{\prime\prime}B^{\prime\prime}}
V_{M^\prime B^\prime,M^{\prime\prime}B^{\prime\prime}}(E)
G_{M^{\prime\prime}B^{\prime\prime}}(E) 
t_{M^{\prime\prime}B^{\prime\prime},MB}(E)\,,
\label{eq:cc-mbmb}
\end{eqnarray}
with
\begin{eqnarray}
V_{M^\prime B^\prime,MB}(E)= v_{M^\prime B^\prime,MB}
+Z_{{M}^\prime {B}^\prime,MB}(E) \,.
\label{eq:veff-mbmb}
\end{eqnarray}
Here $Z_{{M}^\prime {B}^\prime,MB}(E)$
contains the effects due to the coupling with $\pi\pi N$ states.
It has the following form
\begin{eqnarray}
\fl
Z_{{M}^\prime {B}^\prime,MB}(E)
&=& [< {M}^\prime {B}^\prime \mid F
\frac{ P_{\pi \pi N}}
{E- H_0 - \hat{v}_{\pi \pi N}+ i\epsilon}
F^\dagger  \mid {M} {B} > ]_{connected}\,,
\label{eq:z-mbmb0}
\end{eqnarray}
with
\begin{eqnarray}
\hat{v}_{\pi\pi N} &=& v_{\pi N,\pi N}+ v_{\pi\pi} +v_{\pi\pi N,\pi\pi N}
\label{eq:hatv-pipin}\,, \\
F &=& g_V + v_{MB,\pi\pi N}\,,
\label{eq:vertex-f}
\end{eqnarray}
where $g_V$ has been defined in Eq.(\ref{eq:gv}).
Note that the dis-connected term in Eq.(\ref{eq:z-mbmb0})
 is already included in the mass shifts
$\Sigma_{MB}$ of the propagator Eq.(\ref{eq:g-mb-1})
and must be removed to avoid double counting. 

\begin{figure}
\centering
\includegraphics[width=10cm,angle=-0]{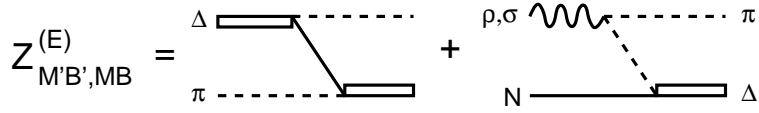}
\caption{One-particle-exchange interactions
 $Z^{(E)}_{\pi\Delta,\pi\Delta}(E)$, $Z^{(E)}_{\pi\Delta,\rho N}$
and $Z^{(E)}_{\pi\Delta,\sigma N}$ of
Eq. (\ref{eq:z-mbmb-e}).}
\label{fig:z}
\end{figure}

The appearance of the projection operator $P_{\pi\pi N}$ in 
Eqs.(\ref{eq:sigma-m}) and (\ref{eq:z-mbmb0}) is the
consequence of the  unitarity condition Eq.(\ref{eq:unitarity}).  
To isolate the effects entirely due to the vertex interaction 
$g_V=\Gamma_{ \Delta\rightarrow \pi N}+
h_{\rho\rightarrow \pi\pi} +h_{\sigma\rightarrow \pi\pi}$, we use
the operator relation
\begin{eqnarray}
\frac{1}{E - H_0 - v}
= \frac{1}{E - H_0} + \frac{1}{E - H_0}v\frac{1}{E - H_0 - v}
\end{eqnarray}
to decompose the $\pi\pi N$ propagator
of Eq.(\ref{eq:z-mbmb0}) to write
\begin{eqnarray}
Z_{M^\prime B^\prime,MB}(E)
= Z^{(E)}_{M^\prime B^\prime,MB}(E)
+ Z^{(I)}_{M^\prime B^\prime,MB}(E) \,.
\label{eq:z-mbmb-s}
\end{eqnarray}
The first term is
\begin{eqnarray}
 Z^{(E)}_{M^\prime B^\prime,MB}(E)
&=& [ < M^\prime B^\prime \mid g_V
\frac{ P_{\pi \pi N}}
{E- H_0  + i\epsilon}  g^\dagger_V
\mid  M B >]_{connected}\,.
\label{eq:z-mbmb-e}
\end{eqnarray}
Obviously, $Z^{(E)}_{M^\prime B^\prime,MB}(E)$ is the one-particle-exchange
interaction between unstable particle channels $\pi\Delta$, $\rho N$, and
$\sigma N$, as illustrated in Fig.\ref{fig:z}.
The second term of Eq.(\ref{eq:z-mbmb-s}) is
\begin{eqnarray}
\fl
 Z^{(I)}_{M^\prime B^\prime,MB}(E)
= < M^\prime B^\prime \mid F
\frac{ P_{\pi \pi N}} {E- H_0 +  i\epsilon}
t_{\pi \pi N,\pi\pi N}(E)
\frac{ P_{\pi \pi N}} {E- H_0 +  i\epsilon}
F^\dagger \mid M B > \nonumber \\
  + <M^\prime B^\prime \mid g_V
\frac{ P_{\pi \pi N}}
{E- H_0  + i\epsilon} v^\dagger_{\pi\pi N,MB} 
\mid M B > \nonumber \\
 + < M^\prime B^\prime \mid v_{M^\prime B^\prime,\pi\pi N}
\frac{ P_{\pi \pi N}}
{E- H_0  + i\epsilon}  g^\dagger_V
\mid  M B> \nonumber \\
  + < M^\prime B^\prime \mid v_{M^\prime B^\prime,\pi\pi N}
\frac{ P_{\pi \pi N}}
{E- H_0  + i\epsilon} v^\dagger_{\pi\pi N,MB} 
\mid M B > \,.
\label{eq:z-mbmb-i}
\end{eqnarray}
Here $t_{\pi\pi N,\pi\pi N}(E)$ is a three-body scattering amplitude
defined by
\begin{eqnarray}
\fl
t_{\pi\pi N, \pi\pi N}(E) = \hat{v}_{\pi\pi N} +
\hat{v}_{\pi\pi N}
\frac{1}{E-K_\pi-K_\pi-K_N-\hat{v}_{\pi\pi N} + i\epsilon} \hat{v}_{\pi\pi N}\,,
\label{eq:tpinpin}
\end{eqnarray}
where $\hat{v}_{\pi\pi N}$ has been defined in Eq.(\ref{eq:hatv-pipin}).

\begin{figure}
\centering
\includegraphics[width=10cm,angle=-0]{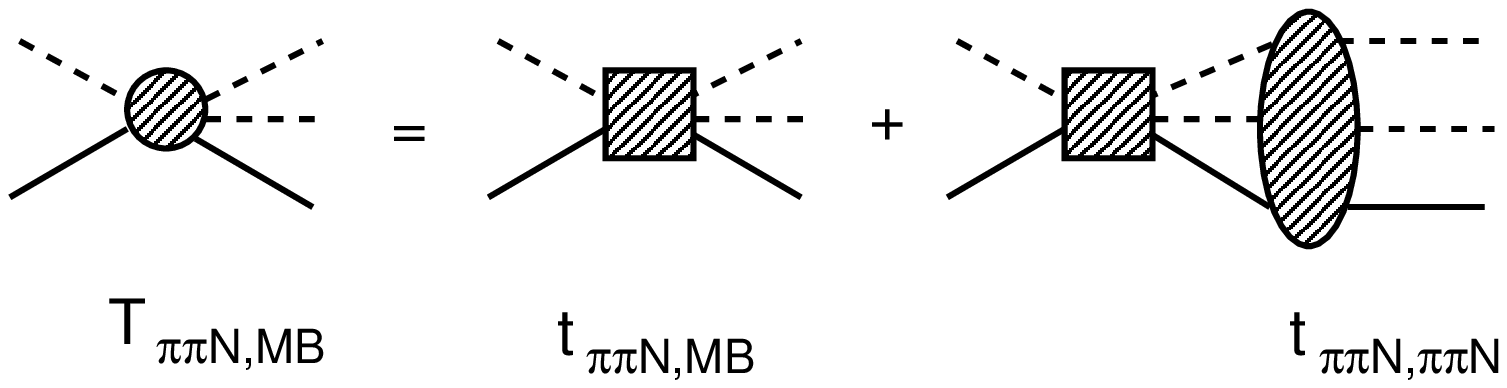}\\
\includegraphics[width=5cm,angle=-90]{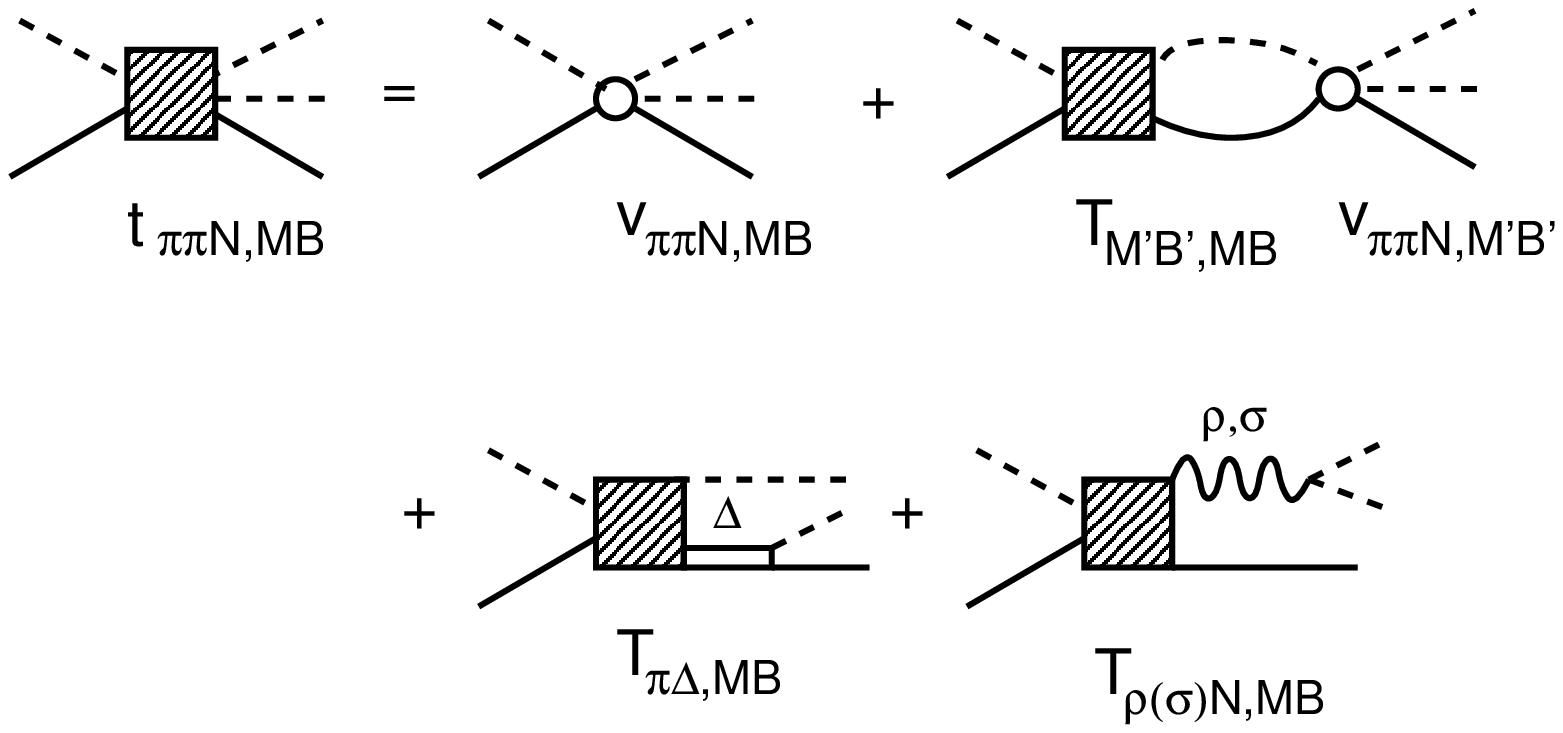}
\caption{Graphical representations of  $T_{\pi\pi N,MB}$
defined by Eqs. (\ref{eq:tpipin-1})-(\ref{eq:omega-pipin}).}
\label{fig:mb-pipin-1}
\end{figure}

The amplitudes $T_{M'B',MB}=t_{M'B',MB}+t^R_{M'B',MB}$ defined by
Eq.(\ref{eq:tmbmb}) can be used directly to calculate the cross sections
of $\pi N \rightarrow \pi N, \eta N$ and 
$\gamma N \rightarrow \pi N, \eta N$ reactions. 
They are also the input to the calculations of the
two-pion production amplitudes.
The two-pion production amplitudes resulted from our derivations 
 are illustrated in 
Fig.\ref{fig:mb-pipin-1}. They can be cast exactly into the following
form
\begin{eqnarray}
\fl
T_{\pi\pi N,MB}(E) = T^{dir}_{\pi\pi N,MB}(E)+
T^{\pi\Delta}_{\pi\pi N,MB}(E)+T^{\rho N}_{\pi\pi N,MB}(E)
+T^{\sigma N}_{\pi\pi N,MB}(E)\,,
\label{eq:tpipin-1}
\end{eqnarray}
with
\begin{eqnarray}
\fl
T^{dir}_{\pi\pi N,MB}(E)
=
<\psi^{(-)}_{\pi\pi N}(E)|
 \sum_{M'B'}v_{\pi\pi N,M'B'}[\delta_{M'B',MB}  \nonumber \\
  +G_{M'B'}(E)(t_{M'B',MB}(E) 
+t^R_{M'B',MB})] | MB>
\label{eq:tpipin-dir} \,, \\
\fl
T^{\pi \Delta}_{\pi\pi N,MB}(E)
=
 <\psi^{(-)}_{\pi\pi N}(E)|
\Gamma^\dagger_{\Delta\rightarrow \pi N}
G_{\pi\Delta}(E)[t_{\pi\Delta, MB}(E)+t^R_{\pi\Delta, MB}(E)] | MB>
\label{eq:tpipin-pid} \,, \\
  \nonumber \\
\fl
T^{\rho N}_{\pi\pi N,MB}(E)
=
 <\psi^{(-)}_{\pi\pi N}(E)|
h^\dagger_{\rho\rightarrow \pi\pi}
G_{\rho N}(E)[t_{\rho N, MB}(E)+t^R_{\rho N, MB}(E)] | MB>
\label{eq:tpipin-rhon} \,, \\
  \nonumber \\
\fl
T^{\sigma N}_{\pi\pi N,MB}(E)
=
 <\psi^{(-)}_{\pi\pi N}(E)|
h^\dagger_{\sigma \rightarrow \pi\pi}
G_{\sigma N}(E)[t_{\sigma N, MB}(E)+t^R_{\sigma N, MB}(E)] | MB> \,.
\label{eq:tpipin-sigman-ext}
\end{eqnarray}
In the above equations, the $\pi\pi N$ scattering wave function is defined by
\begin{eqnarray}
<\psi^{(-)}_{\pi\pi N}(E)|&=&
<\pi\pi N|\Omega^{(-)\dagger}_{\pi\pi N}(E) \,,
\end{eqnarray}
where the scattering operator is defined by
\begin{eqnarray}
\fl
\Omega^{(-)\dagger}_{\pi\pi N}(E)
 =
<\pi\pi N|[1+ t_{\pi\pi N, \pi\pi N}(E)\frac{1}
{E-K_\pi-K_\pi-K_N + i\epsilon}]\,. \label{eq:omega-pipin}
\end{eqnarray}
Here the three-body scattering amplitude
 $t_{\pi\pi N, \pi\pi N}(E)$ is 
 determined by the
non-resonant interactions $v_{\pi\pi}$, $ v_{\pi N,\pi N}$ and 
$v_{\pi\pi N, \pi\pi N}$, as defined by Eq.(\ref{eq:tpinpin}).

We note here that the direct production amplitude
$T^{dir}_{\pi\pi N,MB}(E)$ of Eq.(\ref{eq:tpipin-dir}) is due to 
$v_{\pi\pi N, MB}$ interaction, while
the other three terms are through the unstable $\pi\Delta$, $\rho N$,
and $\sigma N$ states illustrated in Fig.\ref{fig:mb-pipin-1}.
Each term has the contributions from the
non-resonant amplitude $t_{M'B',MB}(E)$
and resonant term $t^R_{M'B',MB}(E)$.

\section{Cross Sections and  $N$-$N^*$ Transition Form Factors}

In this section, we give formula for calculating the cross sections
of all electroweak pion production reactions. Their relations with the
commonly used CGLN and mutipole amplitudes are given in appendix B.
For later discussions in section 5, we also present formula for
calculating the electromagnetic $N$-$N^*$ transition form factors which
are the main focus of recent studies of electromagnetic meson
production reactions.

\subsection{Cross Section Formula}

With the relation Eq.(\ref{eq:smatrix}) between the S- and T- matrices and the normalization
$<\vec{k}|\vec{k}^{\,'}> = \delta (\vec{k}-\vec{k}^{\,'})$, 
the amplitude  $T_{\gamma N,\pi N}$ 
for the pion photoproduction reaction
$\gamma(\vec{q}) + N(-\vec{q}) \rightarrow \pi(\vec{k}) + N(-\vec{k})$ 
defined by Eq. (\ref{eq:tmbmb}) can be written in the final $\pi N$ center of mass
frame as (suppressing spin-isospin indices)
\begin{eqnarray}
T_{\pi N,\gamma N}  & = &
     \frac{1}{(2\pi)^3}\frac{m_N}{\sqrt{E_N(q) E_N(k) 2E_\pi(k) 2q} }
     [e J_{em}\cdot\epsilon_\gamma]. \label{t-gamma-j}
\end{eqnarray}
Here $\epsilon_\gamma$ is the polarization vector of photon.
In the tree-diagram approximation,
 the current matrix element $J_{em}$ is of the form of 
$[\bar{u}_{-\vec{p}}\, I \,u_{-\vec{q}}]$, where $I$ is the usual invariant
amplitudes calculated from the Lagrangian $L(x)=j^\mu_{em}(x) A_\mu(x)$,
where $j^\mu_{em}(x)$ is the electromagnetic current operator and
$A_\mu(x)$ is the electromagnetic field.
Similarly the amplitudes for the electroweak pion production reactions
$e(p_e) + N(p) \rightarrow e'(p_{e'}) + \pi(k) + N(p')$,
 $\nu_e(p_\nu) + N(p) \rightarrow e^-(p_{e'}) + \pi(k) + N(p')$,
and  $\nu(p_\nu) + N(p) \rightarrow \nu(p_{\nu'}) + \pi(k) + N(p')$
can be written as
\begin{eqnarray}
T_{e'\pi N,e N}  & = &
     \frac{1}{(2\pi)^{9/2}}\frac{m_N}{\sqrt{E_N(p) E_N(p') 2E_\pi(k)} }
      \frac{e^2}{q^2}J_{em}\cdot L_{em}, \\
T_{e'\pi N,\nu_e N}  & = &
     \frac{1}{(2\pi)^{9/2}}\frac{m_N}{\sqrt{E_N(p) E_N(p') 2E_\pi(k) }}
      \frac{G_F \cos\theta_c}{\sqrt{2}}J_{cc}\cdot L_{cc},\\
T_{\nu'\pi N,\nu N}  & = &
     \frac{1}{(2\pi)^{9/2}}\frac{m_N}{\sqrt{E_N(p) E_N(p') 2E_\pi(k) }}
      \frac{G_F}{\sqrt{2}}J_{nc}\cdot L_{nc},  \label{t-nunc-j}
\end{eqnarray}
where $J^\mu_{cc}$, $J^\mu_{nc}$ are the matrix elements of  charged current
and neutral current, respectively.
The lepton current matrix elements are 
\begin{eqnarray}
L^\mu_{em} & = & \bar{u}(p_{e'})\gamma^\mu u(p_e),  \\
L^\mu_{cc} & = & \bar{u}(p_{e'})\gamma^\mu(1-\gamma_5)u(p_\nu), \\
L^\mu_{nc} & = & \bar{u}(p_{\nu'})\gamma^\mu(1-\gamma_5)u(p_\nu).
\end{eqnarray}
The pion production current, 
$J^\mu_\alpha$($\alpha=em,cc,nc$)
can be written in terms of  commonly used 
CGLN amplitudes  and
multipole amplitudes. These are summarized in  appendix B.

The differential cross sections of  pion productions
reactions due to electromagnetic ($em$) 
and charged weak current ($cc$) 
 in the massless leptons ($m_e =0$) limit 
can be written as
\begin{eqnarray}
\frac{d\sigma_{em}^5}{d E_{e'} d\Omega_{e'} d\Omega_\pi^*}
=
\frac{1}{4}\frac{e^4}{Q^4}\frac{E_{e'}}{E_e}\frac{Q^2}{1-\epsilon}
 \frac{E k_{\pi}}{2\pi^3 m_N}
  (\frac{m_N}{4\pi E})^2 R_{em}, \label{eq:dcrst-em}\\
\frac{d\sigma_{cc}^5}{d E_{e'} d\Omega_{e'} d\Omega_\pi^*}
=\frac{G_F^2\cos^2\theta_c}{2}\frac{E_{e'}}{E_\nu}
  \frac{Q^2}{1-\epsilon}\frac{E k_{\pi}}{2\pi^3 m_N}
  (\frac{m_N}{4\pi E})^2 R_{cc}. \label{eq:dcrst-cc}
\end{eqnarray}
where $E$ is the invariant mass of the final $\pi N$ 
state,
$\epsilon$ is defined by the lepton scattering angle $\theta_{lep}$
as $\epsilon  =  1/[1 + 2 \frac{|\bm{q}_L|^2}{Q^2}
  \tan^2 \frac{\theta_{lep}}{2}]$ and $k_\pi$ is pion momentum
in the $\pi N$ center of mass system.
The functions $R_{\alpha}$
depends on the pion angle with respect to the direction of momentum
transfer $\vec{q}$ and also the angle $\phi_\pi$ between the
the $\pi-N$ plane and the plane of the incoming and outgoing leptons.
Explicitly, we have
\begin{eqnarray}
R_{em} & = &
  R_{em}^T + \epsilon R_{em}^L
  +\sqrt{2\epsilon(1+\epsilon)}R^{LT}_{em,c} \cos \phi_\pi 
  + \epsilon R^{TT}_{em,c} \cos 2\phi_\pi ], \label{eq:resp-em}\\
R_{cc} & = &
  R_{cc}^T + \epsilon R_{cc}^L
  +\sqrt{2\epsilon(1+\epsilon)}(R_{cc,c}^{LT} \cos \phi_\pi
                              + R_{cc,s}^{LT} \sin \phi_\pi)
\nonumber  \\
 & & + \epsilon(R_{cc,c}^{TT} \cos 2\phi_\pi + R_{cc,s}^{TT} \sin 2\phi_\pi ) \,.
\end{eqnarray}
The structure functions $R^\beta_\alpha$ in the above equations
 are calculated from
the current $J^\mu_{\alpha}$ for the $N + j_\alpha \rightarrow \pi + N$
introduced in Eqs. (\ref{t-gamma-j})-(\ref{t-nunc-j}) in the 
pion-nucleon center of mass system.
It is common to choose the momentum transfer of leptons as the quantization
z-direction $\vec{q}= |\vec{q}\ |(0,0,1)$ and set
the outgoing pion on the x-z plane 
$\vec{k}_\pi =|\vec{k}_\pi|(\sin\theta,0,\cos\theta)$.
The structure functions 
 can then be written as
\begin{eqnarray}
 R_{\alpha}^T & = &   \sum [\frac{|J_{\alpha}^x|^2 + |J_{\alpha}^y|^2}{2}
  - \sqrt{1-\epsilon^2} \mbox{Im}(J_{\alpha}^x J_{\alpha}^{y *})]\,,
\label{eq:resp-t} \\
 R_{\alpha}^L & = & \sum \frac{Q^2}{|\bm{q}_c|^2}|\bar{J}_{\alpha}^0|^2\,,\\
 R^{LT}_{\alpha,c} & = & \sum \sqrt{\frac{Q^2}{|\bm{q}_c|^2}}
{} [  -\mbox{Re}(\bar{J}_{\alpha}^0 J_{\alpha}^{x*}) 
     + \sqrt{\frac{1-\epsilon}{1+\epsilon}}
\mbox{Im}(\bar{J}_{\alpha}^0 J_{\alpha}^{y*}) ]\,,
    \\
 R^{LT}_{\alpha,s} & = & \sum \sqrt{\frac{Q^2}{|\bm{q}_c|^2}}
{}[  \mbox{Re}(\bar{J}_{\alpha}^0 J_{\alpha}^{y*}) 
     + \sqrt{\frac{1-\epsilon}{1+\epsilon}}
\mbox{Im}(\bar{J}_{\alpha}^0 J_{\alpha}^{x*}) ]\,,
  \\
 R^{TT}_{\alpha,c} & = & \sum \frac{|J_{\alpha}^x|^2 - |J_{\alpha}^y|^2}{2}\,,\\
 R^{TT}_{\alpha,s} & = &  - \sum \mbox{Re}(J_{\alpha}^x J_{\alpha}^{y *})\,,\\
\label{eq:resp-1}
\end{eqnarray}
where $\alpha=em,cc$, and we have defined
\begin{eqnarray}
\bar{J}_{\alpha}^0  & =& J_{\alpha}^0 + \omega q\cdot J_{\alpha}/Q^2 \,.
\label{eq:resp-0}
\end{eqnarray}
The spin sum of the nucleons  $\sum$ is
\begin{eqnarray}
\sum = \frac{1}{2}\sum_{s_N,s_N'} \,.
\end{eqnarray}
For investigating the weak pion production reactions induced by
$\mu$ neutrinos, the above
formula need to be modified to include the finite mass $m_\mu$ of the outgoing 
$\mu$ lepton. These formula were given 
in Ref. \cite{Sato:2003rq} and were used in obtaining the  results to be
reviewed in section 6.2.
The cross section formula for the neutral current reactions 
can be obtained  by replacing $G_F\cos\theta_c $ and
$J_{cc}$ of  Eq. (\ref{eq:dcrst-cc}) with $G_F$ and $J_{nc}$.

For the structure functions of the electromagnetic current $R_{em}$, 
we use $\bar{J}^0=J^0$ and 
$\mbox{Im}(J_{em}^xJ_{em}^{y,*})=\mbox{Im}(J_{em}^0J_{em}^{y,*})=0$.
For pion electroproduction cross sections, it is convenient to
write
Eqs.(\ref{eq:dcrst-em}) as
\begin{eqnarray}
\frac{d\sigma_{em}^5}{d E_{e'} d\Omega_{e'} d\Omega_\pi^*}
 & = & \Gamma_T \frac{d\sigma^v}{d\Omega^*_\pi}\,,
\end{eqnarray}
with
\begin{eqnarray}
\Gamma_T & = & \frac{\alpha}{2\pi^2 Q^2}
\frac{E_{e'}}{E_e}\frac{q_{\gamma,L}}{1-\epsilon}\,,\\
\frac{d\sigma^v}{d\Omega^*_\pi} & = &
\frac{k_{\pi}}{q_{\gamma}}
 (\frac{m_N}{4\pi E})^2 e^2 R_{em}\,,
\end{eqnarray}
where $q_{\gamma} = (E^2 - m_N^2)/(2E)$
and $q_{\gamma,L} = (E^2 - m_N^2)/(2m_N)$.

\subsection{$N^*$ Transition Form Factor}

The main objective of analyzing the data of electromagnetic 
meson production reactions is 
to extract the
$\gamma N \rightarrow \bar{N}^*(JT)$ transition form factors
with $J$ and $T$ denoting the spin and isospin of a nucleon resonance.
 In this section, we
define these quantities within  our formulation.

Our starting point is the
following Lagrangian density within the framework of
the relativistic quantum field theory
\begin{eqnarray}
L_{em}(x) = e j^\mu_{em} (x) A_\mu(x)\,, \nonumber
\end{eqnarray}
where $A_\mu(x)$ is the electromagnetic field and $j_{em}^\mu(x)$ is the
current operator.
In the rest frame of $\bar{N}^*$,
the electromagnetic $\gamma N(s_z,t_z) \rightarrow \bar{N}^*(JT)$ transition form 
factors
are usually characterized\cite{Copley:1969ft,Aznauryan:2008us} by the helicity amplitudes $A_\lambda$ for the
spatial components and $S_{1/2}$ for the time component 
of currents :
\begin{eqnarray}
A^{JT}_{3/2,t_z}(Q^2) & = &
X<\bar{N}^*(JT)| \vec{j}_{em}(Q^2)\cdot\vec{\epsilon}_{1}|N(s_z=1/2,t_z)>\,, \label{a32}\\
A^{JT}_{1/2,t_z}(Q^2) & = &
X<\bar{N}^*(JT)|\vec{j}_{em}(Q^2)\cdot\vec{\epsilon}_{1}|N(s_z=-1/2,t_z)>\,,\label{a12}\\
S^{JT}_{1/2,t_z}(Q^2) & = &
X<\bar{N}^*(JT)|j^0_{em}(Q^2)|N(s_z=1/2,t_z)>\,, \label{s12}
\end{eqnarray}
where $Q^2 = -q^2= \vec{q}^{\,2}- \omega^2$ is defined by the photon  momentum
$q^\mu = (\omega, \vec{q})$, and
\begin{eqnarray}
X =   \frac{e}{\sqrt{2K_\gamma}}, \\
\vec{\epsilon}_1  =  \frac{\vec{e}_x+ i \vec{e}_y}{\sqrt{2}}.
\end{eqnarray}
The effective photon energy is determined by the
resonance mass $M_{res}$ 
as $K_\gamma = (M_{res}^2 - m_N^2)/(2M_{res})$.
The helicity amplitudes  Eqs. (\ref{a32})-(\ref{a12}) 
are related to the radiative decay width of the $\bar{N}^*$ as
\begin{eqnarray}
[\mbox{Width}]_{\gamma, t_z}(\bar{N}^*(JT))
=\frac{K_\gamma^2}{4\pi}\frac{m_N}{M_{N^*}}\frac{8}{2J+1}
[|A^{JT}_{3/2,t_z}|^2+|A^{JT}_{1/2,t_z}|^2] \,.
\label{eq:gn-heli}
\end{eqnarray}

\begin{figure}
\centering
\includegraphics[width=8cm,angle=-0]{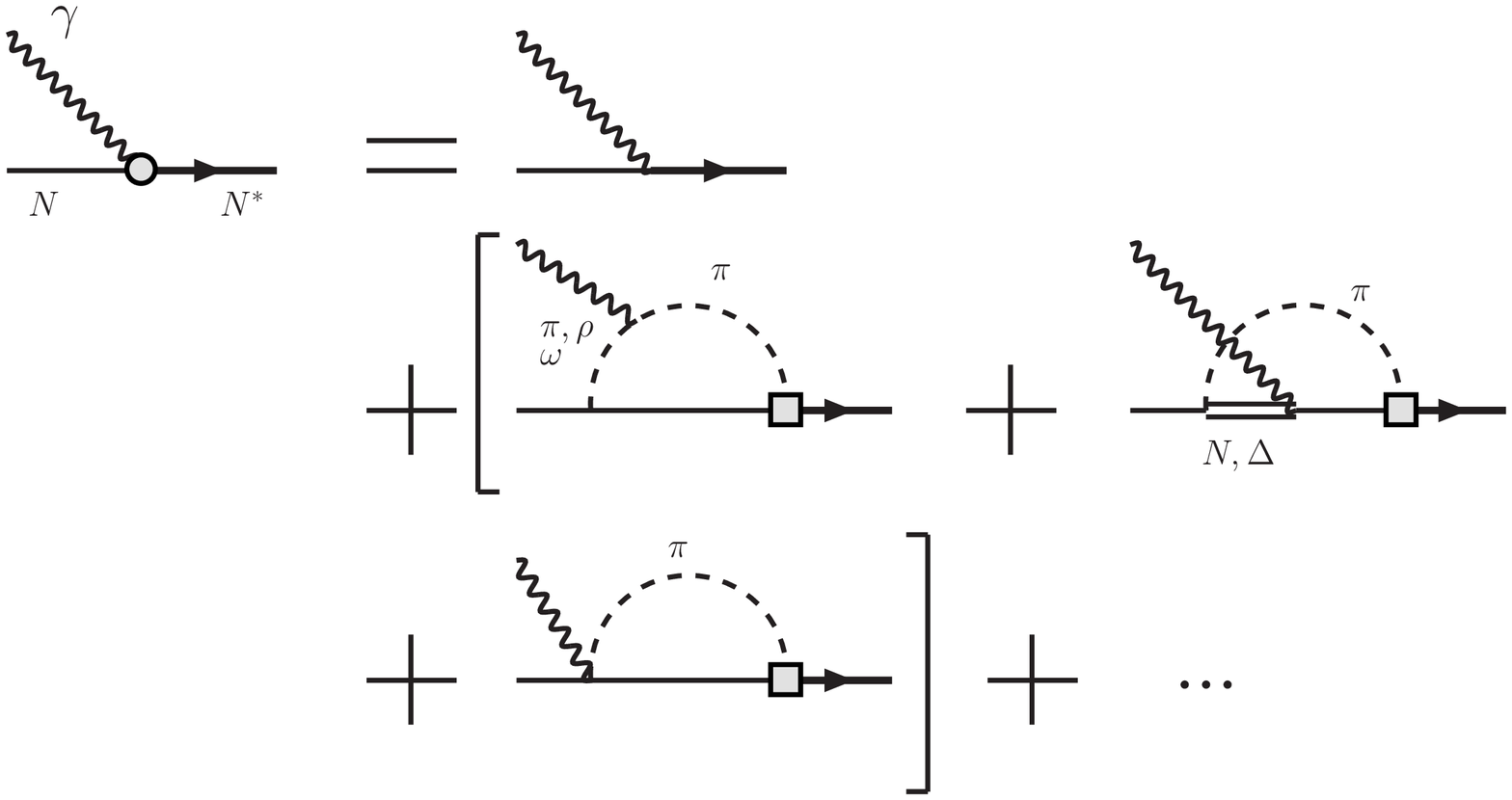}
\caption{Graphical representation of the dressed
$\bar{\Gamma}_{\gamma N \rightarrow N^*}$ defined by 
Eqs.(\ref{eq:ff-dress})-(\ref{eq:ff-mc})}
\label{fig:mb-cloud}
\end{figure}

Since the nucleon resonances couple with the meson-baryon continuum
states, 
the $\bar{N}^*$ state vector appearing in Eqs. (\ref{a32})-(\ref{s12})
is an eigenstate (Gamow state) of the Hamiltonian at the
resonance energy $E_{res}=(M_{res}, -i\Gamma_{res}/2)$ which is
defined by the condition $E_{res} = M^0_{N^*} +  \bar{\Sigma}(E_{res})$.
It consists of  a bare $N^*$ state  and meson-baryon components 
\begin{eqnarray}
\fl
|\bar{N}^*(JT)> &=& |N^*(JT)> \nonumber \\
\fl
& & + 
\sum_{MB,M'B'}(\delta_{MB,M'B'} +
t_{MB,M'B'}G_{M'B'})\Gamma_{N^*\rightarrow M'B'}|N^*(JT)> \nonumber \\
\fl
&=& |N^*(JT)> + \sum_{MB}|MB><MB|\bar{\Gamma}_{N^*\rightarrow MB}|N^*(JT)>. \label{reswf}
\end{eqnarray}
Here we have used the relation Eq.(\ref{eq:nstar-mb}) for defining the
dressed vertex $\bar{\Gamma}_{N^*\rightarrow MB}$.
Thus the form factors defined by Eqs.(\ref{a32})-(\ref{s12})
are determined by the following matrix elements
\begin{eqnarray}
<\bar{N}^*(JT)|j^\mu_{em}|N>\cdot\epsilon_\mu
=<{N}^*(JT)|j^\mu_{em}|N> + \delta_{mc} \,,
\label{eq:ff-dress}
\end{eqnarray}
where the meson cloud effects are
\begin{eqnarray}
\fl
\delta_{mc}=
\sum_{MB} <N^*(JT)|\bar{\Gamma}_{N^*\rightarrow MB}|MB> G_{MB}
[<MB|j_{em}^\mu|N>\cdot \epsilon_\mu]\,.
\label{eq:ff-mc}
\end{eqnarray}
The matrix element $[<MB|j_{em}^\mu|N>\cdot \epsilon_\mu]$ defines
 the non-resonant $v_{MB,\gamma N}$ parts of 
the interaction $v_{22}$ of Eq. (\ref{eq:v22}).
Eq. (\ref{eq:ff-dress}) is illustrated in Fig. \ref{fig:mb-cloud}.

Our normalization is chosen such that the vertex functions 
$\Gamma_{\gamma N \rightarrow N^*}$ and 
$\bar{\Gamma}_{\gamma N \rightarrow N^*} $ of
Eqs.(\ref{eq:mb-nstar})-(\ref{eq:nstar-mb}) 
in each partial wave are related to 
the matrix element of the current 
operator by
\begin{eqnarray}
<N^*(JT)|e j_{em}\cdot \epsilon |N>
  &=&  \sqrt{\frac{2J+1}{4\pi}} \Gamma_{\gamma N \rightarrow N^*}(JT) \,,
 \nonumber \\
<\bar{N}^*(JT)|e j_{em}\cdot \epsilon |N>
  &=&  \sqrt{\frac{2J+1}{4\pi}} \bar{\Gamma}_{\gamma N \rightarrow N^*}(JT)
\,.
\nonumber
\label{current}
\end{eqnarray}

For comparing with theoretical predictions from hadron models and
LQCD, we need to evaluate
the helicity amplitudes Eqs. (\ref{a32})-(\ref{s12})
at the resonance pole $E_{res}$.
This is a non-trivial problem  and is 
being investigated in Ref. \cite{Suzuki:2008rp}.

\section{Results}

With the formulation presented in the above two sections,
very extensive
data of $\pi N$, $\gamma N$, $N(e,e^\prime)$ and also $N(\nu_\mu, \mu \pi)$
reactions have been analyzed.
Most detailed results\cite{Sato:1996gk,Sato:2000jf,Sato:2003rq,Matsui:2005ns}
 are for the $\Delta (1232)$ state. These will be reviewed in
subsection 6.1 for the electromagnetic
$\gamma N\rightarrow \pi N$ and $N(e,e'\pi)$ processes and  6.2 for
the weak $N(\nu_\mu, ,\mu\pi)$ reactions.
The investigation
of higher mass $N^*$ states began in 2006 and is still
in the progressing stage. Thus only limited results will be reviewed in
subsection 6.3.

\subsection{Electromagnetic Excitation of the $\Delta (1232)$ state}

The electromagnetic excitation of
the $\Delta (1232)$ state was studied in  Refs.
\cite{Sato:1996gk,Sato:2000jf,JuliaDiaz:2006xt}.
The main objective was to extract the $\gamma N \rightarrow \Delta (1232)$ form
factors from the data of photoproduction and electroproduction of $\pi$
in the invariant mass $W \leq $ 1.3 GeV region where only $\pi N$
and $\gamma N$ channels are open. Thus it was studied by using the 
formula presented in section 4 by keeping only one bare $\Delta$ state and
including only the $\pi N$ and $\gamma N$ channels. The resulting model
is identical to the model developed in Refs.\cite{Sato:1996gk} (called
the  Sato-Lee (SL) model in the literatures).

\begin{figure}[tb]
\vspace{10pt}
\begin{center}
\mbox{\epsfig{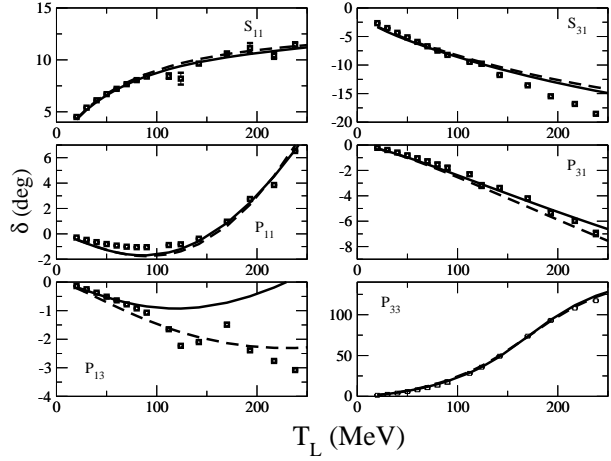}}
\end{center}
\caption{Phase shifts of $\pi N$ elastic scattering up to $T_L=$250 MeV.
Solid and dotted stand for model SL and SL2 respectively. Data, $L_{2T,2J}$,
are from the energy independent SAID~\protect\cite{Arndt:2003fj} analysis plus 8 points from
their energy dependent solution for the $P_{13}$ and $P_{31}$ partial
waves at lower energies.
\label{fig:pshift}}
\end{figure}

The $\gamma N \rightarrow \Delta$ (1232) form factor
 $\Gamma_{\Delta,\gamma N}$ is parametrized
in the form developed by Jones and Scadron~\cite{Jones:1972ky}. With the
normalization $<{\vec k}|{\vec k'}> = \delta({\vec k}-{\vec k'})$
 for the plane wave 
states and $<\phi_B | \phi_{B'}> = \delta_{B,B'}$ for 
$B = N $ and bare $\Delta$ states, the covariant form of
Jones and Scadron can be cast, in the rest frame of
the $\Delta$ and for the photon momentum $q=(\omega,\vec {q})$,
as 
\begin{eqnarray}
\fl
& & <m_{j_\Delta},  m_{t_\Delta} 
| \Gamma_{\Delta,\gamma N}(q)| \lambda_\gamma\lambda_N, m_{t_N}> \nonumber \\ 
\fl
& & =  F \times \langle \thalf m_{t_\Delta} |\ohalf  1 m_{t_N} 0\rangle 
\nonumber \\
\fl
& & \,\,\,\,\,\times [ M_{m_{j_\Delta},\lambda_\gamma\lambda_N}(q) G_M(Q^2)
+E_{m_{j_\Delta},\lambda_\gamma\lambda_N}(q)G_E(Q^2) 
 +C_{m_{j_\Delta},\lambda_\gamma\lambda_N}(q)G_C(Q^2)]\,,
\label{eq:mxf}
\end{eqnarray}
where $<jm|j_1,j_2,m_1,m_2>$ is the Clebsch-Gordon coefficient
of $\vec{j}_1 + \vec{j}_2=\vec{j}$ coupling, $\lambda_\gamma$ and $\lambda_N$ are the helicities of the
initial photon and nucleon, $m_{j_\Delta}$ is the z-component of
the $\Delta$ spin, $m_{t_\Delta}$ and $m_{t_N}$ denote the isospin
components. In Eq.(\ref{eq:mxf}) we have defined   
\begin{eqnarray}
F=\frac{-e}{(2\pi)^{3/2}}\sqrt{\frac{E_N(\vec{q})+m_N}{2E_N(\vec{ q})}}
\frac{1}{\sqrt{2\omega}}\frac{3(m_\Delta+m_N)}{4m_N(E_N(\vec{q})+m_N)}\,,
\end{eqnarray}
and the excitation kinematics are contained in 
\begin{eqnarray}
M_{m_{j_\Delta},\lambda_\gamma\lambda_N}(q) &=&
<m_{j_\Delta}|i\vec{ S}\times \vec{ q}\cdot \vec{ \epsilon}_{\lambda_\gamma}
|\lambda_N>\,,
\label{eq:coef-m}\\
E_{m_{j_\Delta},\lambda_\gamma\lambda_N}(q) &=&
<m_{j_\Delta}|\vec{ S}\cdot \vec{ \epsilon}_{\lambda_\gamma}
\vec{ \sigma}\cdot \vec{q}
+\vec{ S}\cdot \vec{ q}\vec{ \sigma}\cdot\vec{ \epsilon}_{\lambda_\gamma}|\lambda_N>\,,
 \label{eq:coef-e} \\
C_{m_{j_\Delta},\lambda_\gamma\lambda_N}(q) &=&\frac{1}{m_\Delta}
<m_{j_\Delta}|\vec{S}\cdot\vec{ q}\vec{ \sigma}\cdot\vec{ q}\epsilon_0|
\lambda_N>\,,
 \label{eq:coef-c}
\end{eqnarray}
where $e=\sqrt{4\pi/137}$, photon polarization vector
is defined by
$\vec{\epsilon}_{\pm 1}=\mp \frac{1}{\sqrt{2}}(\hat{x}\pm i \hat{y})$, and
$\epsilon^0_{\pm 1} =  0 $ for $\lambda_\gamma =\pm 1$, 
$\vec{\epsilon}_0=0$  and  $\epsilon^0_0 =  1 $
for the scalar component $\lambda_\gamma =0 $.
The transition spin $\vec{S}$ is defined
by $<j_\Delta m_\Delta |S_m|j_Nm_N>=<j_\Delta m_\Delta|j_N 1 m_N m>$.

The form factors $G_M(Q^2)$, $G_E(Q^2)$, and $G_C(Q^2)$
in Eq.(\ref{eq:mxf}) describe magnetic
M1, Electric E2, and Coulomb C2 transitions.  
Choosing the photon direction $\vec{ q}$ in the z-direction,
the above form factors 
are related to the form factors in helicity representation 
defined in Eqs.(\ref{a32})-(\ref{s12}), which
are consistent with the convention of Particle Data Group~\cite{PDG:2007} (PDG)
\begin{eqnarray}
A_{3/2}(Q^2) &=& - \frac{\sqrt{3}A}{2} [G_M(Q^2) +  G_E(Q^2)] \,,
\label{eq:he-1a} \\
A_{1/2}(Q^2) &=& - \frac{A}{2}[ G_M(Q^2) - 3 G_E(Q^2)]  \,,
\label{eq:he-2a}\\
S_{1/2}(Q^2)&=& - \frac{|\vec{q}|A}{\sqrt{2}m_\Delta} G_C(Q^2) \,,
\label{eq:he-3a}
\end{eqnarray}
with
\begin{eqnarray}
 A & = & \frac{e}{2m_N}\sqrt{\frac{m_\Delta }{m_N K_\gamma}}
       \frac{|\vec{q}|}{1 + Q^2/(m_N+m_\Delta)^2}\,,
\end{eqnarray}
where $K_\gamma = \frac{m_\Delta^2 - m_N^2}{2m_\Delta}$.

The dressed form factor $\bar{\Gamma}_{\Delta,\gamma N}$
has the same symmetry property of the bare vertex defined above. Thus it
can be expanded in the same form of Eq.(\ref{eq:mxf}).
We denote the dressed form factors by $\bar{G}_M(Q^2)$, $\bar{G}_E(Q^2)$,
$\bar{G}_C(Q^2)$. The corresponding helicity amplitudes $\bar{A}_\lambda$
can also be calculated by 
using the same relations Eqs.(\ref{eq:he-1a})-(\ref{eq:he-3a}).
In Ref.~\cite{Sato:1996gk}, 
it was shown that $\bar{\Gamma}_{\Delta,\gamma N}$ 
can also be calculated from the K-matrix form of 
$\bar{\Gamma}_{\Delta,\gamma N}$ which is
 directly related to the imaginary parts of
the $full$ multipole amplitudes $M_{1+}$, $E_{1+}$ and
$S_{1+}$ at the
resonance energy $W_R$ where the $\pi N$ phase shift is $90^0$, independent
of the form of the non-resonant amplitudes.
Thus the  
dressed ratios can 
be calculated from
\begin{eqnarray}
\bar{R}_{EM} (W= W_R)&=& - \frac{\bar{G}_E}{\bar{G}_M}= \frac{{\rm Im} E_{1+}}{{\rm Im} M_{1+}}\,, \\
\bar{R}_{SM} (W=W_R) &=&  \frac{|\vec{q}|}{2m_\Delta} \frac{\bar{G}_C}{\bar{G}_M}= \frac{{\rm Im} S_{1+}}{{\rm Im} M_{1+}}\,,
\end{eqnarray}
It is common to define $G^*_M$ for the M1 transition form factor
which is related to our dressed form factor by
\begin{eqnarray}
G^*_M(Q^2)= \sqrt{\frac{\Gamma^{exp}_\Delta}
{\Gamma^{SL}_\Delta}}\frac{\bar{G}_M(Q^2)}{\sqrt{1+Q^2/(m_\Delta+m_N)^2}}\,,
\label{eq:gmstar}
\end{eqnarray}
where $\Gamma^{exp}_\Delta = 115$ MeV is used in extracting the data from
$M^{3/2}_{1+}$ amplitude of pion electroproduction amplitude and
$\Gamma^{SL}=93 $ MeV from the constructed model.

\begin{figure}[p]
\vspace{10pt}
\begin{center}
\mbox{\epsfig{file=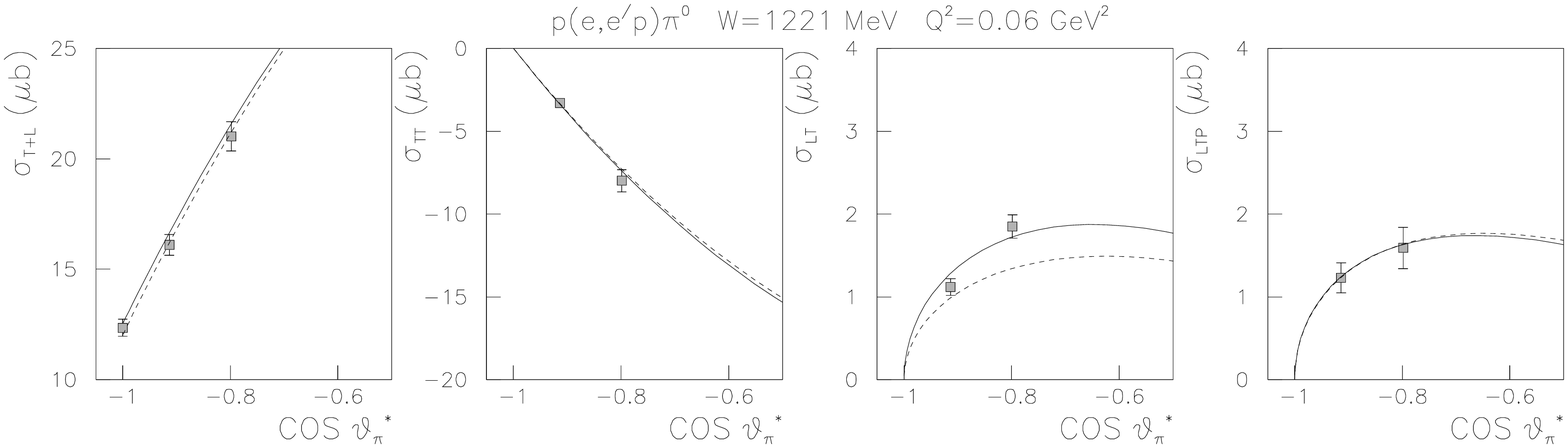, width=12cm}}
\mbox{\epsfig{file=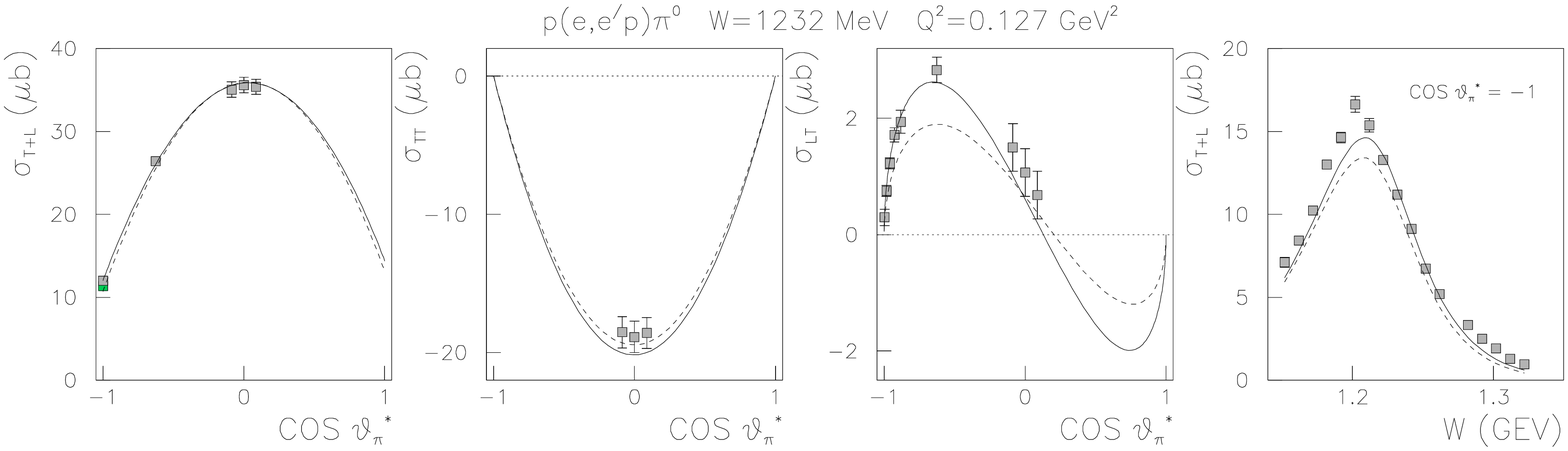, width=12cm}}
\mbox{\epsfig{file=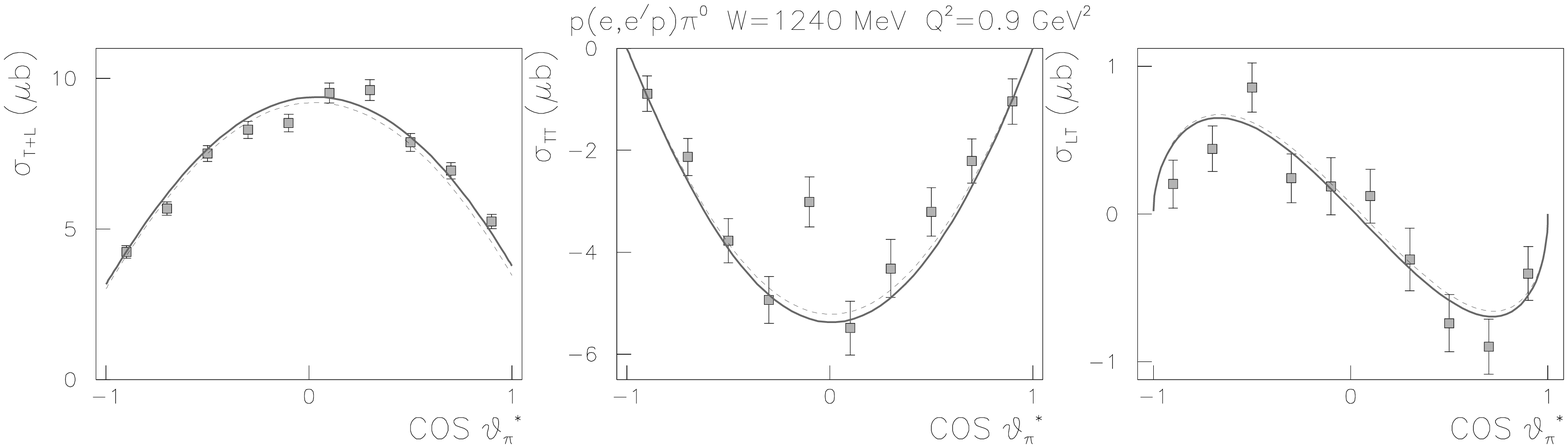, width=12cm}}
\mbox{\epsfig{file=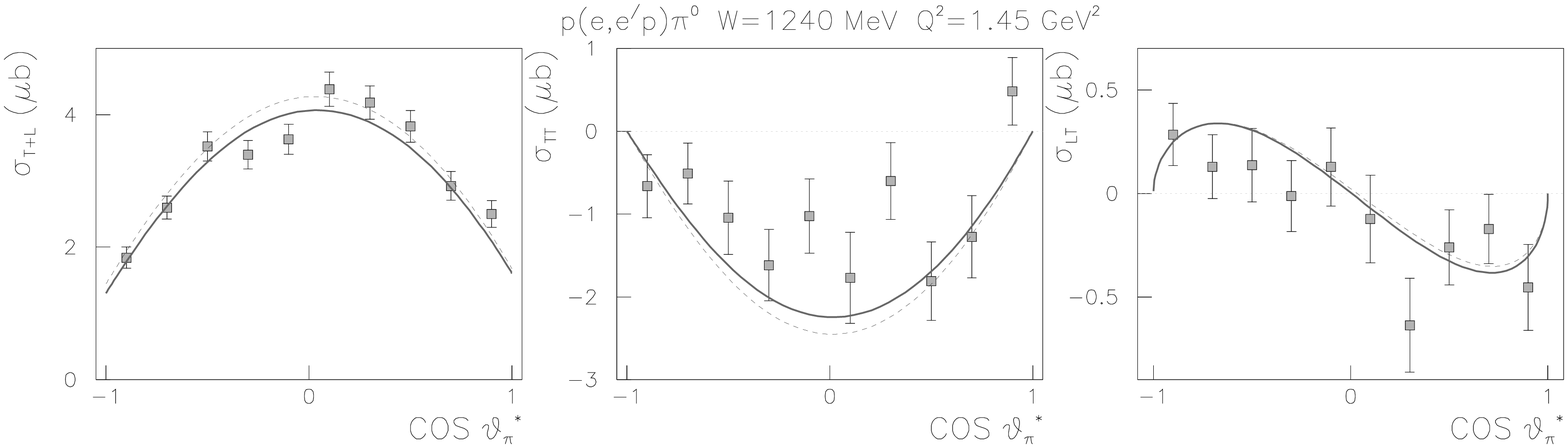, width=12cm}}
\end{center}
\caption{Fits to experimental $p(e,e^{\prime}p)\pi^0$
structure functions. Solid lines are from the fits with
the bare form factors $G_M(Q^2)$, $G_E(Q^2)$ and $G_C(Q^2)$
adjusted at each $Q^2$. The dashed curves are from
the calculations using the parametrization 
Eqs.(\ref{sl-gm0})-(\ref{eq:r-sl}).
The structure functions $\sigma_{\alpha}$ are $R^{\alpha}_{em}$
defined in Eq.(\ref{eq:resp-em}).
Data are from MAMI~\protect\cite{Stave:2006ea} at $Q^2=0.06$~GeV$^2$, 
BATES~\protect\cite{Mertz:1999hp,Kunz:2003we,Sparveris:2004jn}
at $Q^2=0.127$~GeV$^2$, CLAS~\protect\cite{Smith:2006} ($W=1220$~MeV) and 
MAMI~\protect\cite{Sparveris:2006} ($W=1221$~MeV)
at $Q^2=0.2$~GeV$^2$ and CLAS~\protect\cite{Joo:2001tw} at $Q^2=0.9,1.45$~GeV$^2$.}
\label{fig:slfit1}
\end{figure}

\begin{figure}[t!]
\vspace{10pt}
\begin{center}
\mbox{\epsfig{file=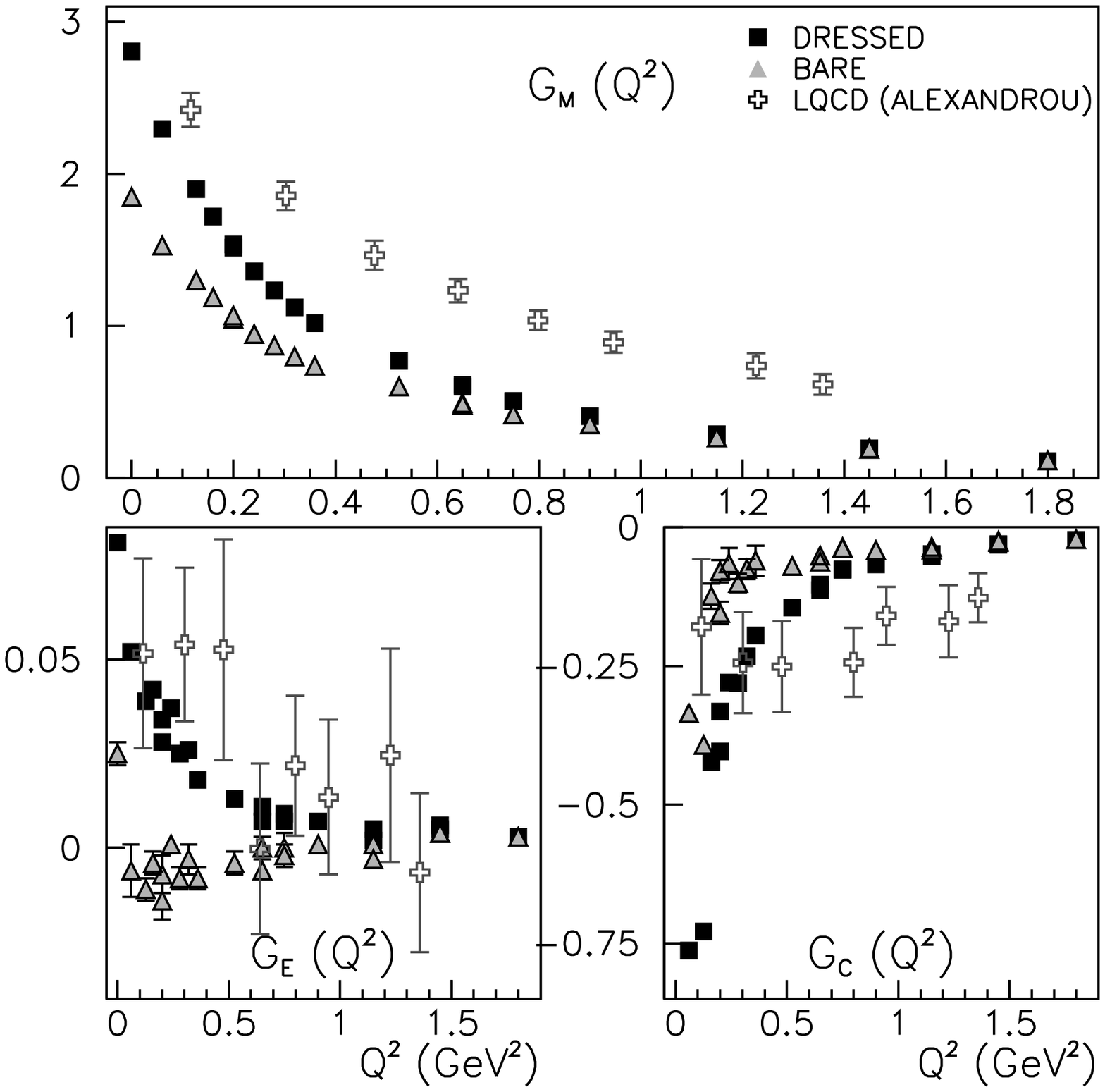, width=8cm}}
\end{center}
\caption{
The extracted $\gamma N \rightarrow \Delta$ form factors.
Dark squares(triangles) are the dressed (bare) values.
Open crosses with errors are the lattice QCD calculation of
Ref.~\protect\cite{Alexandrou:2003ea,Alexandrou:2004xn}.}
\label{fig:lqcd}
\end{figure}

With the above definitions of $\gamma N \rightarrow \Delta$ (1232)
form factors, we now describe the results obtained 
in Refs.\cite{Sato:1996gk,Sato:2000jf,JuliaDiaz:2006xt}.
The first step in extracting the $\gamma N\rightarrow \Delta$ (1232) 
form factors is to fix the hadronic parameters by fitting the
$\pi N$ elastic scattering up to $W= $ 1.3 GeV.
Two fits from 
Refs. \cite{Sato:1996gk,JuliaDiaz:2006xt} are shown in Fig.\ref{fig:pshift}.
These two models will be called SL and SL2 models in later discussions.
Their differences are
 mainly in fitting the weak $P_{13}$ partial waves.
These two fits provide us with an opportunity to examine the model 
dependence of the extracted 
$\gamma N \rightarrow \Delta$ (1232) form factors.

The next step is to adjust the bare $\gamma N \rightarrow \Delta$ (1232)
form factors $G_M(Q^2)$, $G_E(Q^2)$,
and $G_C(Q^2)$ to fit the world data of $\gamma p\rightarrow \pi^0 p, \pi^+n$
, $p(e,e^\prime \pi^0)p$ and $p(e,e^\prime \pi^+)n$. In Fig.\ref{fig:slfit1},
we show some typical fits to the structure functions of $p(e,e'\pi^0)p$ .
The resulting bare (solid triangles) and dressed (solid squares)
form factors are shown in Fig.\ref{fig:lqcd}.
In the same figure we also show
the LQCD  results (open crosses with errors) which are obtained from applying 
a chiral extrapolation procedure to get results in the physical region
from the calculations  with very large quark masses.
We see that LQCD results agree only very qualitatively with
either the extracted dressed or bare form factors. There are several
difficulties in interpreting these results, as discussed
by Pascalutsa
and Vanderhaeghen~\cite{Pascalutsa:2005vq}.
First, the chiral extrapolation is only valid for low $Q^2$, 
although it has been used
in a rather high $Q^2$ region. Second, there are higher order corrections
on the commonly used chiral extrapolation, which have not been under
control.
Thus it is not clear what to conclude
from Fig.~\ref{fig:lqcd} for the results from LQCD of 
Ref.~\cite{Alexandrou:2003ea,Alexandrou:2004xn}.
Further investigations are clearly needed.

\begin{table}
\begin{tabular}{|c|cccc|cccc|}
\hline
  & \multicolumn{4}{|c|}{$\bar{R}_{EM}$(\%)} &
\multicolumn{4}{|c|}{$\bar{R}_{SM}$(\%)} \\
  $Q^2$ &  & UIM & SL & SL2 &  & UIM & SL & SL2 \\
\hline
0.16 &  & -1.94(0.13) & -2.45(0.2) & -2.57(0.2) & & -4.64(0.19) & -4.44(0.35) & -4.36(0.35) \\
0.20 &  & -1.68(0.18) & -2.21(0.2) & -2.31(0.2) &  & -4.62(0.18) & -4.23(0.35) & -4.14(0.35) \\
0.24 &  & -2.14(0.14) & -2.70(0.2) & -2.76(0.2) &  & -4.60(0.28) & -4.32(0.35) & -4.21(0.35) \\
0.28 &  & -1.69(0.27) & -1.99(0.2) & -2.07(0.2) &  & -5.50(0.31) & -5.08(0.35) & -4.97(0.35) \\
0.32 &  & -1.59(0.17) & -2.29(0.2) & -2.35(0.2) &  & -5.71(0.33) & -4.87(0.35) & -4.75(0.35) \\
0.36 &  & -1.52(0.27) & -1.80(0.2) & -1.82(0.2) &  & -5.79(0.43) & -4.76(0.35) & -4.56(0.35) \\
\hline
\end{tabular}\caption{Extracted values of $E2/M1$ ratio
$\bar{R}_{EM}$ and $C2/M1$ ratio
$\bar{R}_{SM}=S_{1+}/M_{1+}$ at $Q^2=0.16-0.36$~GeV$^2$
from analysis of results from a CLAS
 measurement~\cite{Smith:2006}
 of the $p(e,e^{\prime}p)\pi^0$ reaction.  Methods used are
 Unitary Isobar Model (UIM) and the SL and SL2 which use hadronic
parameters determined 
in Ref.\cite{Sato:1996gk} and Ref.\cite{JuliaDiaz:2006xt}, respectively. 
  Errors are statistical only.}
\label{tab:1}
\end{table}

\begin{figure}[t]
\centering
\mbox{\epsfig{file=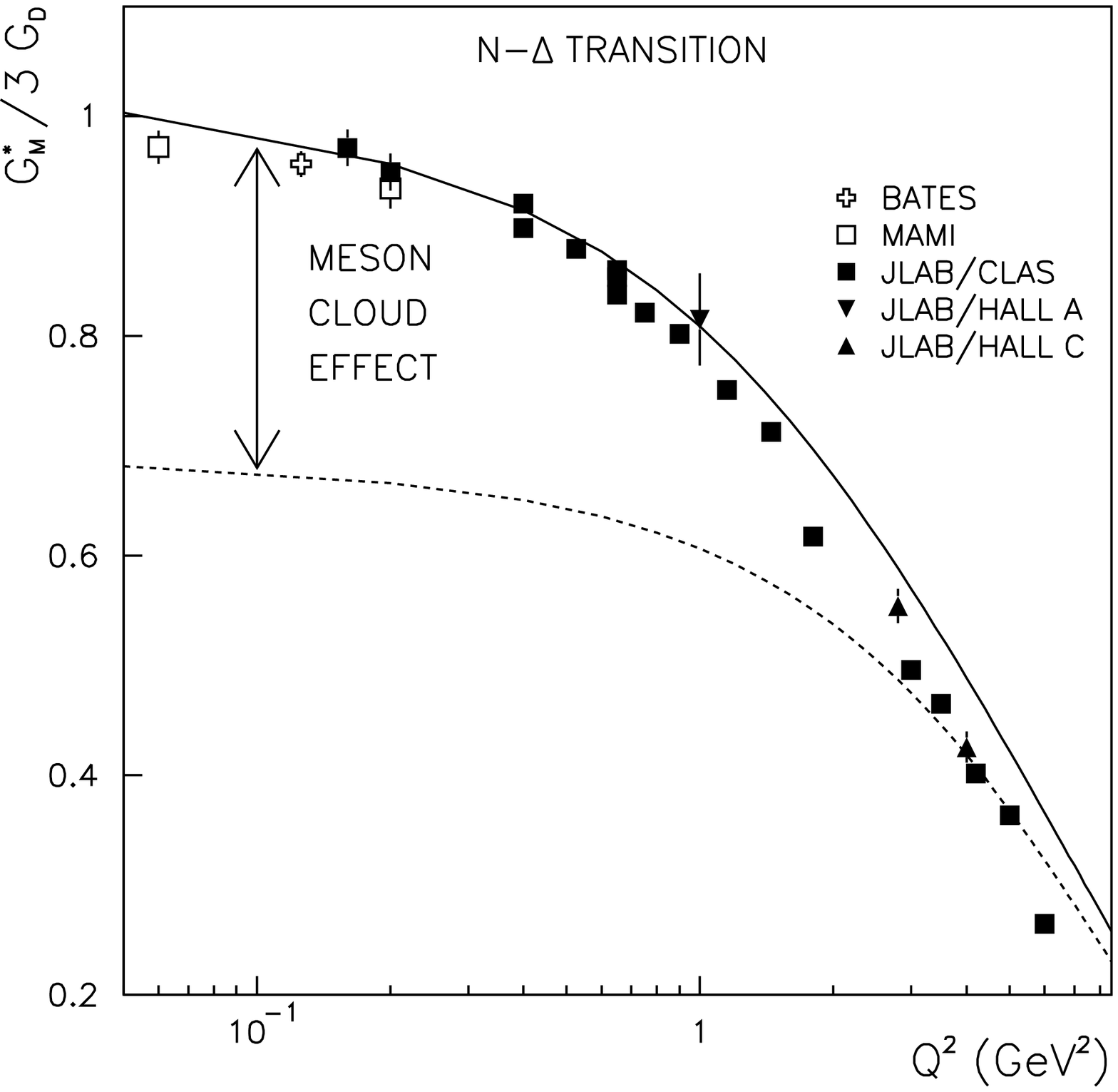, width=8cm}}
\caption{Magnetic dipole transition form factor $G^*_M$
for $\gamma^* N \rightarrow \Delta(1232)$, normalized to
the proton dipole form factor $G_D(Q^2)=1/[1+Q^2/\Lambda^2]^2$ with
$\Lambda^2 = 0.71$ (GeV/c)$^2$.  Experimental points are analyses 
of inclusive data ($\bigcirc$) from pre-1990 experiments at DESY and 
SLAC~\cite{Bartel:1968tw,Adler:1972,Stein:1975,Stuart:1996zs} and recent 
exclusive $p(e,e^{\prime}p)\pi^o$ data (blacksquare) from 
BATES~\cite{Mertz:1999hp,Kunz:2003we,Sparveris:2004jn}, 
MAMI~\cite{Stave:2006ea,Sparveris:2006} and 
JLAB~\cite{Smith:2006,Joo:2001tw,Kelly:2005zc,Kelly:2005,Frolov:1998pw,Ungaro:2006df}.
Solid curve is from the dressed 
calculation of SL model  using the parametrization of 
Eq.~(\ref{eq:r-sl}). The dotted curve
is obtained when the meson cloud effect, defined by Eq.(\ref{eq:ff-mc})
 is turned off.
\label{fig:gm_Delta}}
\end{figure}

Here we note that the extracted bare form factor $G_M(Q^2)$ (solid triangles)
in Fig.~\ref{fig:lqcd} are close to
the following parametrization of Ref.\cite{Sato:2000jf}
\begin{equation}
G_M(Q^2)=G_M(0)R_{SL}(Q^2)G_p(Q^2)\,,
\label{sl-gm0}
\end{equation}
where
$G_p(Q^2)=1/(1+Q^2/M_V^2)^2$ with $M_V^2 = 0.71$ (GeV/c)$^2$ being the
well determined nucleon form factor, and
\begin{eqnarray}
R_{SL}(Q^2)= (1+a\,Q^2)\,exp(-b\,Q^2)\,, 
\label{eq:r-sl}
\end{eqnarray}
with $G_M(0)=1.85$, $a=0.154$ (GeV)$^{-2}$ and $b=0.166$ (GeV)$^{-2}$. 
By using this parametrization, the predicted bare (dotted curve)
and dressed (solid curve) $G^*_M(Q^2)$ ( defined by Eq.(\ref{eq:gmstar})) 
are
compared with the available empirical values in Fig.\ref{fig:gm_Delta}.
It is clear that the resulting dressed $G^*_M(Q^2)$ (solid curve)
 agree well with the available empirical values.
The differences between the solid and dotted curves indicate
that the meson cloud effects, illustrated in Fig.\ref{fig:mb-cloud},
 are important
in the  low $Q^2$ region and gradually diminish as $Q^2$ increases.
This result is one of the main accomplishments of
many-year study of $N$-$\Delta$ (1232) excitation, and
has motivated  future studies up to $Q^2= 11$ (GeV)$^2$
with 12 GeV upgrade of CEBAF at JLab.  

Historically, the $\Delta$ (1232) is described by the constituent quark model.
To see the extent to which the extracted $G_M(Q^2)$ form factors can be
understood with this model, 
it is instructive to first consider the naive
s-wave non-relativistic quark model within which 
$\mu_p$ for the proton magnetic moment
and $\mu_{\Delta^+p}$ for the $\Delta^+$-$p$ M1 transition are defined by
\begin{eqnarray}
\frac{e}{2m_p}\mu_p = \langle p, m_{s_N}
= \ohalf|\sum_{i}\frac{e_i}{2m_q}\sigma_i(z)|p, m_{s_N}=\ohalf\rangle\,, \\
\frac{e}{2m_p} \mu_{\Delta^+ p} 
= \langle\Delta^+, m_{s_\Delta}
=\ohalf |\sum_{i}\frac{e_i}{2m_q}\sigma_i(z)|p, m_{s_N}=\ohalf\rangle \,.
\end{eqnarray}
From the above relation and the definition Eq.(\ref{eq:mxf}),
one observes that the magnetic M1
form factor of $\gamma N \rightarrow \Delta$
 at $Q^2=0$ can be directly calculated from the
proton magnetic moment
\begin{eqnarray}
\fl
G_{M}(0)&=& [\sqrt{2}G_p(0)]\left[\frac{2(E_N(q)+m_N)}{3(m_\Delta+m_N)}\right]
\sqrt{\frac{2E_N(q)}{E_N(q)+m_N}} = 0.84 \mu_p\,.
\end{eqnarray}
where $q=(m^2_\Delta-m^2_N)/2m_\Delta \sim 260$ MeV/c. If we use
the empirical value of proton magnetic moment 
$\mu_p \rightarrow \mu^{exp}_p = 1+\kappa_p \sim 2.77$, we then find
$G_M(0) \sim 2.32 $ which is considerably smaller than the extracted
dressed value $\sim 3.2$ seen in Fig.~\ref{fig:lqcd}.
This was observed in Ref.\cite{Sato:1996gk} and interpreted as due to
the large meson cloud effects which are the difference
between the solid and dotted curves in Fig.\ref{fig:gm_Delta}.

We thus observe that extracted bare value $G_M(0)=1.85$ can perhaps
be understood
in terms of constituent quark degrees of freedom if we tune properly
the constituent quark model calculations. 
On the other hand, our extracted bare E2 transition form factor
$G_E(0)$ cannot be understood within the
non-relativistic constituent quark 
model. With the tensor force within the conventional one-gluon-exchange, the 
estimated E2 transition of $\gamma N \rightarrow \Delta$
 is known to be negligibly small compared with the value
calculated from our value $G_E(0) = -0.025$. 
In Ref.\cite{JuliaDiaz:2006xt}, the extracted form factors are also compared
 with relativistic constituent quark models.
Only qualitative agreement is obtained.

We next present 
our determined dressed $\bar{R}_{EM}$ and 
$\bar{R}_{SM}$ in the low $Q^2$ region where very large meson 
cloud effects have been identified in Fig.~\ref{fig:lqcd}. Our 
results, SL and SL2, are listed in table~\ref{tab:1} and compared with the 
values determined using the unitary isobar model (UIM). The 
differences between our values and that from the UIM reflect some 
model-dependence in the extraction.
Here we note that only the data of
five of  the eleven $N(e,e^\prime \pi)N$ independent observables were available
and used in the fits. Thus the differences between
different models  shown in Table \ref{tab:1} are surprisingly small.
So far there is no satisfactory theoretical understanding of the
results of $\bar{R}_{EM}$ and $\bar{R}_{SM}$  shown in Table 1.

\begin{figure}[b]
\centering
\includegraphics[width=5cm]{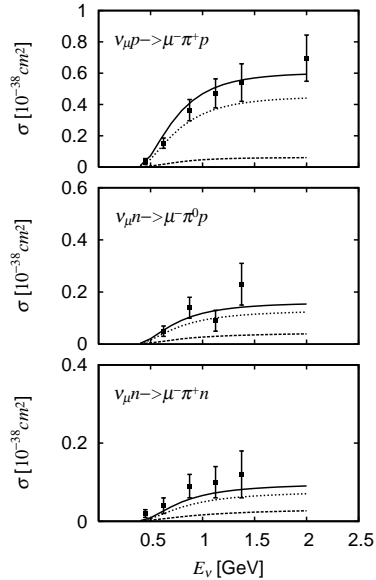}
\caption{Total cross sections of $N(\nu_\mu,\mu^-\pi)N$
reactions predicted by the SL Model\cite{Sato:2003rq}.
The data are from Ref.\cite{Barish:1978pj}. The solid curves are from
full calculations. The dotted curves are from turning off pion cloud
effects on N-$\Delta$ transitions.
The dashed curves are the contributions
from the non-resonant amplitude.  }
\label{fig:neu-tot}
\end{figure}

\subsection{Weak excitation of the $\Delta$ state}

The model developed in Refs.\cite{Sato:1996gk,Sato:2000jf}, the SL model, was extended to
investigate neutrino-induced pion production reactions.
The extension is tedious but straightforward, as detailed in 
Refs.\cite{Sato:2003rq,Matsui:2005ns}. Here we just focus on the 
extraction of the weak $N$-$\Delta$ (1232)
form factor which has  vector ($V$) and
axial vector ($A$) components. 
The vector current matrix element
$<\Delta \mid V^\mu \mid N >$ can be obtained from the SL model by
appropriate isospin rotations. 
The most general
form of the axial vector current matrix element 
is well known, as given in 
Refs.\cite{Adler:1968tw,Adler:1975mt,Hemmert:1994ky}.
To see how it is different from the electromagnetic excitation
given in Eqs.(\ref{eq:mxf}) - (\ref{eq:coef-e}), we cast\cite{Sato:2003rq} it 
 in 
the rest frame of a $\Delta$ on the resonance energy(
 $p_\Delta =  (m_\Delta, \vec{0}),  p_N   =  (E_N(q),-\vec{q}),
 q = (m_\Delta-E_N(q), \vec{q})) $ as
\begin{eqnarray}
<\Delta|\vec{A}^i |N> & = & 
\sqrt{\frac{E_N+m_N}{2m_N}}[
(d_1 + \frac{m_\Delta^2-m_N^2}{m_N^2}d_2)\vec{S} \nonumber \\
& & - (d_2 + d_3)\frac{(\vec{S}\cdot\vec{q})\vec{q}}{m_N^2} 
 -id_4 \frac{\vec{S}\times \vec{q}(\vec{\sigma}\cdot\vec{q})}
         {m_N^2(E_N+m_N)}] T^i \label{eq:d-axial1},\\
<\Delta|A^{0 i} |N> & = & 
\sqrt{\frac{E_N+m_N}{2m_N}}[
 d_2 \frac{\vec{S}\cdot\vec{q}(m_\Delta+E_N)}{m_N^2} \nonumber \\
& & -d_3  \frac{\vec{S}\cdot\vec{q}(m_\Delta-E_N)}{m_N^2}]T^i \label{eq:d-axial2},
\end{eqnarray}
where $T^i$ is the $i-$th component of the
isospin transition operator(defined by the reduced matrix element
$<3/2 \mid\mid {\vec T} \mid \mid 1/2> =
-<1/2 \mid\mid {\vec T}^+ \mid\mid 3/2 > = 2$ in Edmonds convention\cite{Edmond:note}), and the transition spin 
$\vec{S}$ is defined by the same reduced matrix elements of ${\vec T}$.
The above expression suggests that
$d_1,d_2$ terms describe the Gamow-Teller transition 
and $d_4$ describes the quadrupole transition.
For simplicity, we follow Ref.\cite{Hemmert:1994ky} to fix the form factors $d_i(q^2)$ at $q^2=0$
 using the non-relativistic constituent quark model. The axial vector
current  operator for a
constituent quark is derived from taking the
non-relativistic limit of  the standard form
$g_{Aq} \bar{q}\gamma^\mu \gamma_5\frac{\tau}{2} q$. By some
derivations\cite{Sato:2003rq}, we find that
\begin{eqnarray}
d_1(Q^2_0)&=&g^*_A(Q^2_0) ( 1 +\frac{m^2_\Delta - m^2_N}{2m_N(m_\Delta + m_N)})\,, \\
d_2(Q^2_0)&=& - g^*_A(Q^2_0)\frac{m_N}{2(m_\Delta + m_N)}\,, \\
d_3(Q^2_0)&=&-g^*_A(Q^2_0)\frac{m^2_N}{q^2-m_\pi^2}\,,
\end{eqnarray}
where $g_A^*(Q^2_0)=\frac{1}{\sqrt{2}}\frac{6}{5}g_A$ with $g_A =1.26$
 and $Q_0^2=(m_\Delta - m_N)^2$.
This agrees with the results of Ref.\cite{Hemmert:1994ky} if we neglect the  
difference between $m_N$ and $m_\Delta$.

To account for the $q^2$-dependence, we assume that
\begin{eqnarray}
d_i(Q^2)=d_i(0)R_{SL}(Q^2)G_A(Q^2)\,,
\label{eq:ga0}
\end{eqnarray}
where $R_{SL}(Q^2)$ is defined in Eq.(\ref{eq:r-sl}) and has been determined
in the study of $\gamma N \rightarrow \Delta$ (1232) form factor, and
$G_A(Q^2)=1/(1+Q^2/M^2_A)^2$
with $M_A=1.02$ GeV is the nucleon axial form factor\cite{Bernard:2001rs}.

With the axial form factors defined above,
our calculations of $p(\nu_\mu, \mu \pi )N$ do not involve any
adjustment of the parameters, since
all of the the parameters of the non-resonant amplitudes and
the vector part of the $N$-$\Delta$ transition form factor 
have been completely fixed in
the study of electromagnetic pion production.
The predicted total cross sections  
are compared with with the
data\cite{Barish:1978pj} in Fig.\ref{fig:neu-tot}. 
We see that the predictions(solid curves)  agree reasonably
well with the data for three pion channels.
 For the data on neutron target, our predictions(solid curves in 
the middle and lower figures)
are in general lower than the data. This is perhaps related to the 
procedures used in Ref.\cite{Barish:1978pj}
to extract these data from the experiments on
deuteron target. 

Similar to the electromagnetic $N$-$\Delta$ transition, we have also
found significant meson cloud effects on the axial $N$-$\Delta$
transition form factor. This is also shown 
 in  Fig.\ref{fig:neu-tot}.
We see that our full calculations(solid curves) are
reduced significantly to dotted curves
if we turn off the dynamical pion cloud effects.
If we further turn off the contributions 
from bare N-$\Delta$ transitions, 
we obtain the dashed curves which correspond to
the contributions from the non-resonant amplitudes.
Clearly, the non-resonant amplitudes are weaker, but
are also essential in getting the good agreement with the data since
they can interfere with the resonant amplitudes.

\begin{figure}[ht!]
\centering
\includegraphics[width=5cm]{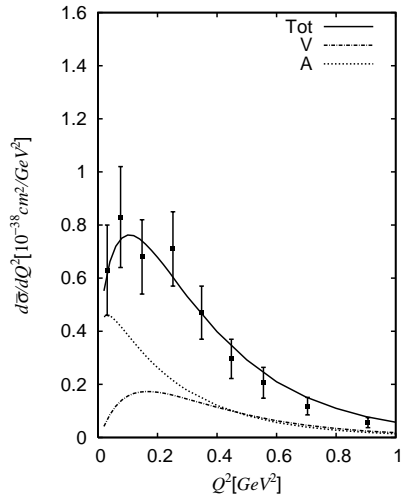}
\caption{Differential cross sections $d\bar{\sigma}/dQ^2$
of  $p(\nu_\mu,\mu^-\pi^+)p$ reaction averaged over
neutrino energies 0.5 GeV $ < E_\nu < $ 6 GeV.
The curves are the predictions of the SL Model\cite{Sato:2003rq}.
The dotted
curve(dot-dashed curve) is the contribution from axial vector
current A (vector current V). The solid curve is from the full
calculations with V-A current.
The data are from Ref.\cite{Barish:1978pj}.}
\label{fig:neut-q2-1}
\end{figure}

In Fig.\ref{fig:neut-q2-1} we compare the $Q^2$-dependence of the
differential cross sections 
$d\sigma/dQ^2$  with the data from ANL\cite{Barish:1978pj}.
We see that our predictions(solid curve) agree reasonable
well with the data both in magnitude and $Q^2-$dependence.
In Fig.\ref{fig:neut-q2-1} we also compare the contributions
from vector current(dot-dashed curve) and axial vector 
current(dotted curve). They have rather different $Q^2$-dependence
in the low $Q^2$ region and interfere constructively with
each other to yield the solid curve of the full results. 
Since vector current contributions are very much constrained by
the $(e,e'\pi)$ data, the results of Fig.\ref{fig:neut-q2-1}
 suggest that the constructed
axial vector currents
are consistent with the data.

\begin{figure}[ht!]
\centering
\includegraphics[width=6cm]{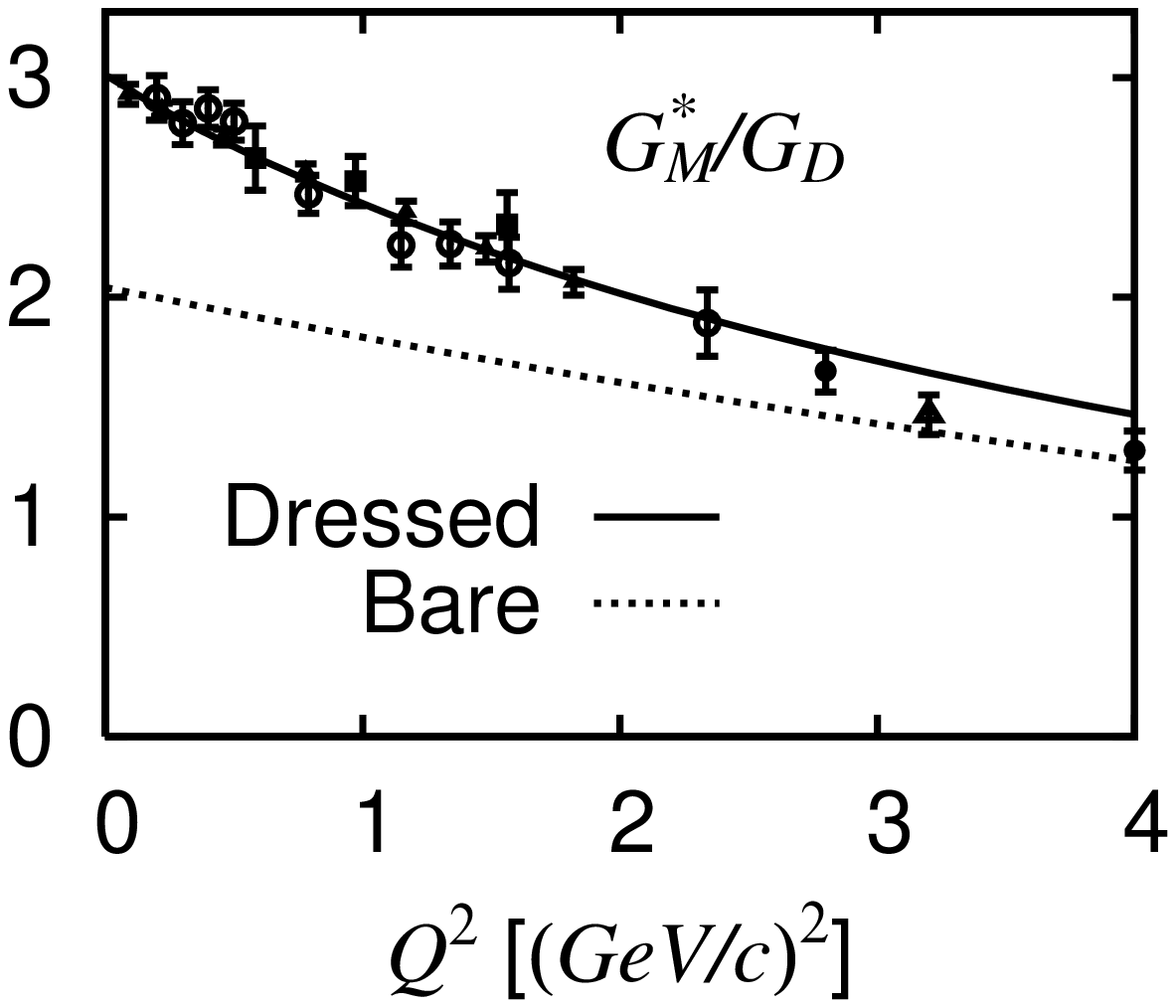}
\includegraphics[width=6cm]{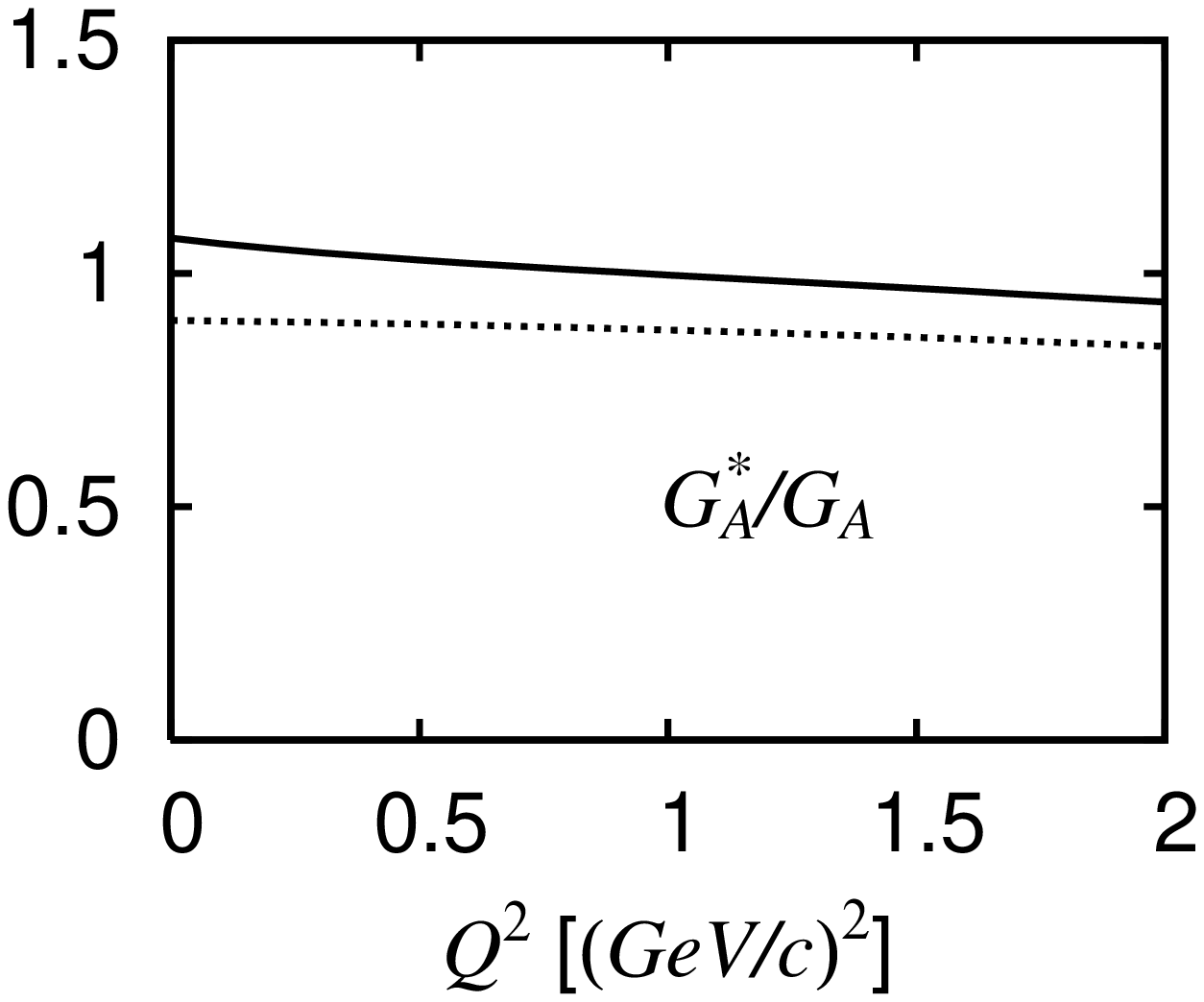}
\caption{The N-$\Delta$ form factors: 
left panel: Magnetic M1 form factors given in Ref.\cite{Sato:2000jf}, 
right panel: axial vector form factor determined in Ref.\cite{Sato:2003rq}. 
The solid curves are from full calculations.
The dotted curves are obtained from turning off the pion cloud
effects. $G_D=1/(1+Q^2/M^2_V)^2$  with $M_V=0.84$ GeV
is the usual proton dipole form factor and $G_A=1/(1+Q^2/M^2_A)^2$ with $M_A=1.02$
GeV is the axial nucleon form factor of Ref. \cite{Bernard:2001rs}.}
\label{fig:gma}
\end{figure}

The extraction of the axial $N$-$\Delta$ form factor is much more difficult
because the lack of sufficient data. 
The dressed (solid curve) and bare (dotted curve)
axial $N$-$\Delta$ form factors are shown in the right-hand side of
Fig.\ref{fig:gma}. Clearly, their  $Q^2$-dependence is weaker than
the $\gamma N \rightarrow \Delta$ form factors which are discussed
in the previous subsection and also displayed in left hand side of
Fig.\ref{fig:gma}. However, the meson cloud effects, the difference
between the solid and dotted curves, are comparable in both form factors.

\begin{figure}[ht!]
\centering
\includegraphics[width=7cm]{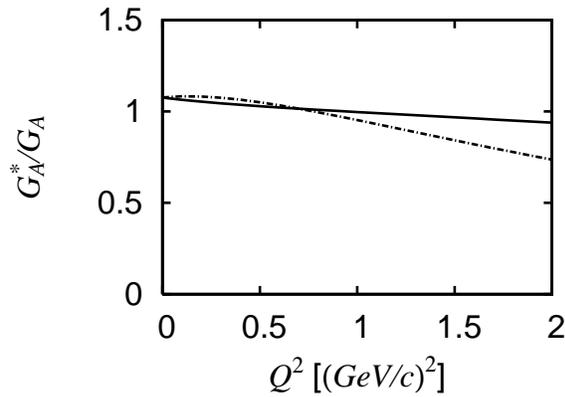}
\caption{Compare the dressed axial N-$\Delta$ form factor predicted by
the Model of Ref.\cite{Sato:2003rq} (solid curve) 
with the empirical form factor(dot-dash curve)
determined in Ref.\cite{Kitagaki:1990vs}.}
\label{fig:gma-exp}
\end{figure}

The  axial $N$-$\Delta$ form factor was determined in previous analysis.
In Fig.\ref{fig:gma-exp}, we see that our results (solid) are significantly
different from the previous results( dot-dashed curve) at high $Q^2$.
Obviously, more experimental data are needed to resolve the differences.
With the new world-wide effort in developing next-generation
neutrino experiments, progress in this direction is expected
in the near future.

\subsection{Excitations of  higher mass $N^*$ states}

To investigate higher mass $N^*$ states up to invariant mass $W=2 $ GeV, 
we apply the full model
developed in sections 3 and 4. 
The meson-baryon (MB) channels considered are 
$ \gamma N, \pi N, \eta N$ and the $\pi\pi N$ channel which
has resonant $\pi\Delta, \sigma N, \rho N$ components. The resonant
amplitude $t^R_{M'B',MB}$ of Eq.(\ref{eq:tmbmb}) 
are generated by including one or two bare $N^*$ states in each partial waves.
Clearly it is a highly nontrivial task to extract the resonance
parameters from solving this multi-channels multi-resonance problem.
It requires  simultaneous fits to all available data of $\pi N$,
$\gamma N$, and $N(e,e')$ data with all possible two-particle and
three-particle $\pi\pi N$ states. 
This ambitious work started in 2006 at the  Excited Baryon 
Analysis Center (EBAC) of JLab, and is still progressing rapidly.
Thus the results reviewed in this subsection are only the first-step
results which will be refined when 
all of the world's meson production
data of $\pi N$, $\gamma N$, and $N(e,e')$ reactions
are included in the analysis. 

\subsubsection{$\pi N$ scattering} 

\begin{figure}[tbp]
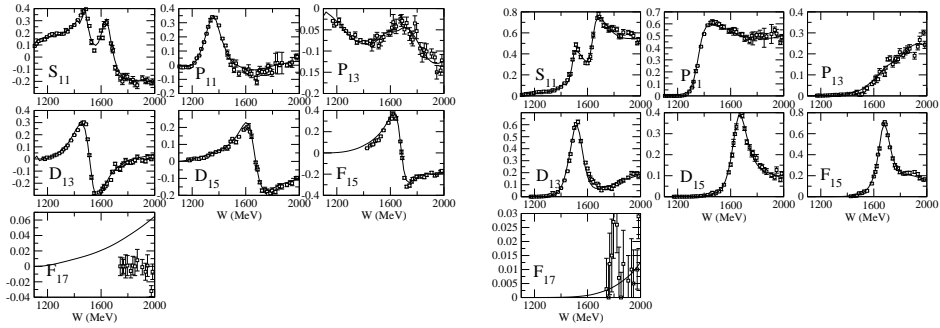

\vspace{20pt}
\begin{center}
\mbox{\epsfig{file=pin-fit1.eps, width=6cm}}\hspace*{0.3cm}
\mbox{\epsfig{file=pin-fit2.eps, width=6cm}}
\end{center}
\caption{The  $\pi N$ partial wave amplitudes
 of isospin $T=1/2$ calculated from the JLMS model\cite{JuliaDiaz:2007kz}
 are compared with
the energy independent solutions of Ref.~\cite{Arndt:2003fj}.\label{figrealiso1}
Left (right) panel is for real (imaginary) parts of the amplitudes}
\label{fig:pin-pwa}
\end{figure}

\begin{figure}
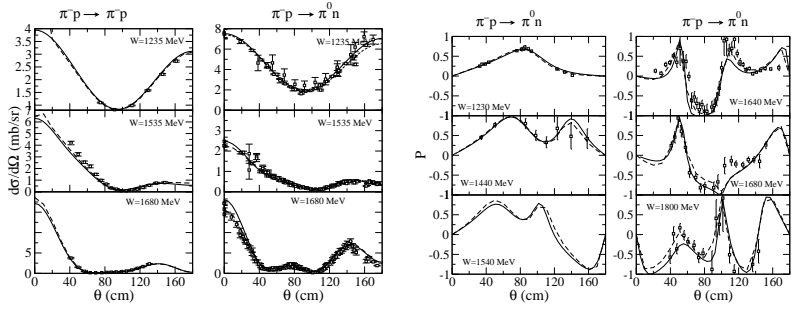

\centering
\includegraphics[width=5cm,angle=-0]{pin-sig-2.eps}\hspace*{0.3cm}
\includegraphics[width=5cm,angle=-0]{pin-p-2.eps}
\caption{Differential cross sections $d\sigma/d\Omega$ (left) and
asymmetry 
$P$ (right) of $\pi^- p \rightarrow \pi^- p, \pi^0 n$ reactions. The solid curves
are from JLMS model\cite{JuliaDiaz:2007kz}. }
\label{fig:pin-crst}
\end{figure}

\begin{figure}[hbt]
\centering
\includegraphics[clip,width=8cm]{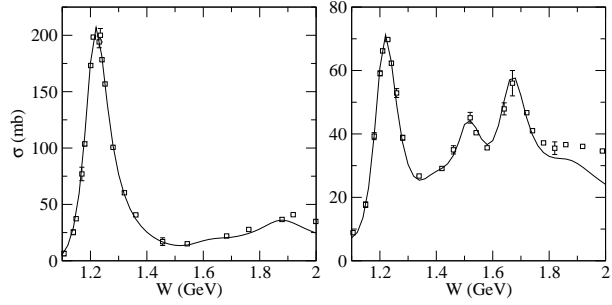}
\caption{Total cross sections of $\pi^+ p$ (left) and $\pi^- p$ (right)
reactions.
Solid curves are from the JLMS model\cite{JuliaDiaz:2007kz}.
Only few data are shown for a clear comparison between curves.
The data are from Refs.~\cite{PDG:2007,CNS:web}.  }
\label{fig:pin-tot}
\end{figure}

Similar to the study of the $\Delta$ (1232) state, the
 first step to investigate higher mass $N^*$ states
is to determine the hadronic parameters by fitting the data of
$\pi N$ elastic scattering. 
 Such a fit was obtained in Ref.\cite{JuliaDiaz:2007kz} by assuming one or
two bare $N^*$ states in each of $S$, $P$, $D$, and $F$ partial waves .
The $\pi N$ scattering amplitudes of isospin $T=1/2$  predicted 
by the resulting model, the JLMS model, are compared 
with the empirical values of SAID\cite{Arndt:2003fj} in Fig.\ref{fig:pin-pwa}.
Similar good agreement is also found for the $T=3/2$ partial waves, as
also given in  Ref.\cite{JuliaDiaz:2007kz}. The corresponding good agreement with
the data of differential cross sections and polarization observable $P$
are illustrated in Fig.\ref{fig:pin-crst} for some of the data.
The predicted total cross sections are also in good agreement 
with the data as shown in Fig.\ref{fig:pin-tot}.

The resulting parameters of 21
bare $N^*$ states, presented in Ref.\cite{JuliaDiaz:2007kz}, is the
starting point for performing a dynamical coupled-channel analysis
of the world's meson production
data of $\pi N$, $\gamma N$, and $N(e,e')$ reactions. In the next three
subsections, we review the results obtained so far. Here we also mention
that it is necessary to develop an analytic continuation method
to identify  the nucleon
resonances with the poles of the scattering amplitudes
on complex energy plane. This has been developed\cite{Suzuki:2008rp}, but will
not be discussed here because of its technical complexities.

\subsubsection{$\pi N \rightarrow \pi\pi N$ reactions}

The main difficulty in fitting the $\pi N$ elastic scattering data, described
above, is that
the model contains many parameters mainly
due to the lack of sound theoretical guidance in parametrizing the bare
$N^* \rightarrow \pi N, \eta N, \pi\Delta, \rho N, \sigma N$ form factors.
Thus  it is necessary to examine
these $N^*$ parameters; in particular the parameters associated with
the unstable $\pi\Delta$, $\rho N$, and $\sigma N$ channels. This has been
done in Ref.\cite{Kamano:2008gr}
in the study of
$\pi N \rightarrow \pi\pi N$ reactions which are known to be dominated
by these unstable particle channels.

\begin{figure}[tb]
\centering
\includegraphics[clip,width=8cm]{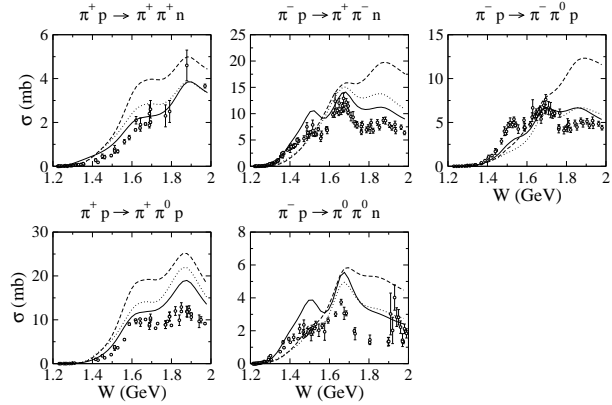}
\caption{The coupled-channels effects on $\pi N \rightarrow \pi\pi N$
reactions. The solid curves are from full calculations, the dotted curves
are from keeping
only $M'B'= MB$ in the Eq.(\ref{eq:no-cc}) and
in $\bar{\Gamma}_{N^* \rightarrow MB}$ of
Eqs.(\ref{eq:mb-nstar})-(\ref{eq:nstar-mb}),
the dashed curves are from setting $t_{MB,M'B'}=v_{MB,M'B'}$.
The data are from~\cite{Arndt:2001by}. }
\label{fig:pinpipin-tot}
\end{figure}

Before we present the predicted $\pi N \rightarrow \pi\pi N$
cross sections, we note here that
the main feature of our approach is a dynamical
 coupled-channels treatment of the unstable
$\pi\Delta, \rho N, \sigma N$ channels.
This effect can be explicitly seen by writing
the coupled-channels equations, Eq.(\ref{eq:cc-mbmb}), as
\begin{eqnarray}
t_{MB,\pi N}(E) = \sum_{M^{'}B^{'}}[1-vG]^{-1}_{MB,M^{'}B^{'}}
v_{M^{'}B^{'},\pi N}\,,
\label{eq:no-cc}
\end{eqnarray}
where $MB=\pi\Delta, \rho N, \sigma N$, and
the intermediate meson-baryon states can be
$M^{'}B^{'}=\pi N,\eta N, \pi\Delta, \sigma N, \rho N$.
The predicted $\pi N \rightarrow \pi\pi N$ total cross sections
depend on the coupled-channel effects due to
 these intermediate $M'B'$ states,

The results for $\pi N \rightarrow \pi\pi N$ total cross sections
 are shown in Fig.\ref{fig:pinpipin-tot}.
We see that our full calculations (solid
curves) can reproduce the data to a very large extent for all
possible $\pi\pi N$ final states up to $W=2$ GeV. These results
are far more successful
than all of the previous investigations, as discussed in Ref.\cite{Kamano:2008gr}.
 When only
the term with $M'B'= MB$ in the Eq.(\ref{eq:no-cc}) and
in $\bar{\Gamma}_{N^* \rightarrow MB}$ of
Eqs.(\ref{eq:mb-nstar})-(\ref{eq:nstar-mb}) is kept, the
calculated total cross sections (solid curves) are changed to the
dotted curves in Fig.\ref{fig:pinpipin-tot}.
 If we further neglect the coupled-channels
effects
by setting $t_{\pi N, MB}=v_{\pi N, MB}$, we then get the
dashed curves which are very different 
from the full calculations (solid curves),
in particular in the high $W$ region.
Clearly coupled-channel effects are very large.

The results shown in Fig.\ref{fig:pinpipin-tot} indicate that
the $N^* \rightarrow \pi \Delta, \rho N, \sigma N$ determined from
fitting $\pi N$ elastic scattering data are reasonable, but clearly need
to be improved. To make the progress in this direction, it is necessary to
have more complete data of $\pi N \rightarrow \pi\pi N$ reactions from
new hadron facilities such as J-PARC. Hopefully, this can be realized
in the near future. At the present time, we  have to
rely on recent data of $\gamma N \rightarrow \pi\pi N$ to refine
the  $N^* \rightarrow \pi \Delta, \rho N, \sigma N$ parameters.
Effort in this direction is being made at EBAC.

\subsubsection{Electromagnetic pion production reactions}
\begin{figure}[t]
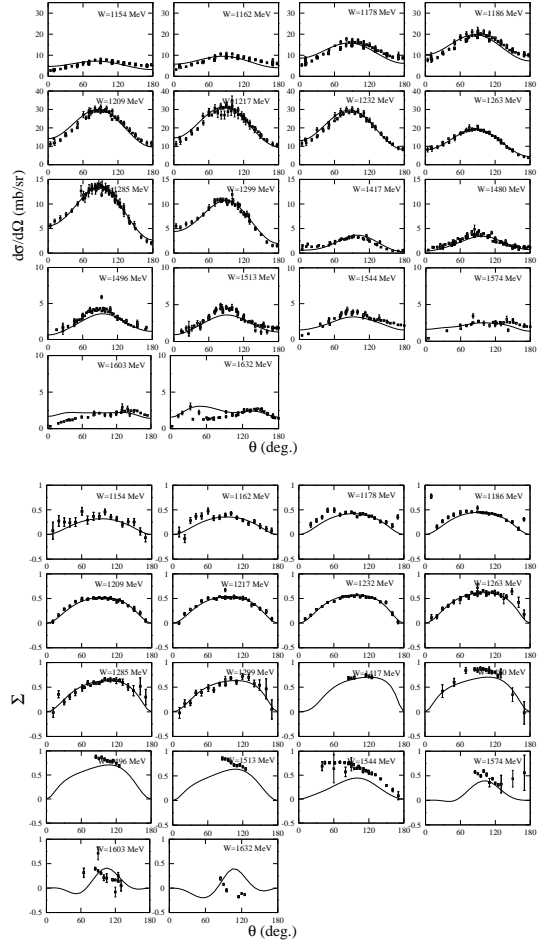

\centering
\includegraphics[width=7cm,angle=-0]{gampi0-sig.eps}\\
\vspace*{0.3cm}
\includegraphics[width=7cm,angle=-0]{gampi0-ay.eps}
\caption{Differential cross section $d\sigma/d\Omega$ (upper) and
photon asymmetry $\Sigma_\gamma$ (lower) for $\gamma p \rightarrow \pi^0 p$ 
calculated from JLMSS model\cite{JuliaDiaz:2007fa} are compared
to experimental data obtained from Ref.~\cite{Arndt:2003fj}.}
\label{fig:gampi0-sig}
\end{figure}

The fits to $\pi N$ reaction data, presented in the previous
two subsections, have fixed all of the hadronic  parameters
of the effective Hamiltonian Eqs.(\ref{eq:H-1})-(\ref{eq:v22}).
Most of the electromagetic parameters associated with the nonresonant 
$\gamma N \rightarrow \pi N$ are also known from previous
investigation of $\Delta$ (1232) state.
Thus the bare helicity amplitudes, $A_{3/2}$, $A_{1/2}$,
 and $S_{1/2}$, defined in Eqs.(\ref{a32})-(\ref{s12}),
are the main unknown parameters in our investigations
of electromagnetic pion production reactions.
The first step in determining these helicity amplitudes
 had been completed
in Ref.\cite{JuliaDiaz:2007fa} by performing
$\chi^2-$fits to the available photoproduction
data of $\gamma N \rightarrow \pi N$
reactions up to $W=1.65 $ GeV.
 The quality of the resulting 
fit can be seen in Figs.~\ref{fig:gampi0-sig} for
the $\gamma p \rightarrow \pi^0 p$. Similar good agreement
was also obtained for the $\gamma p \rightarrow \pi^+ n$,
as also presented in Ref.\cite{JuliaDiaz:2007fa}.  

\begin{figure}[b]
\vspace*{0.2cm}
\centering
\includegraphics[width=10cm,angle=-0]{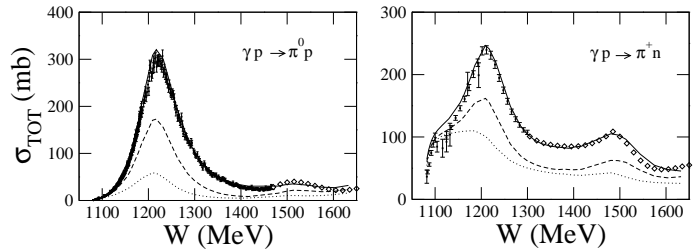}
\caption{Total cross sections from JLMSS model\cite{JuliaDiaz:2007fa}. The dashed curves are obtained from
turning off all $MB$ channels except the $\pi N$
channel in the the loop integrations in the non-resonant
amplitude  and the dressed
$\gamma N \rightarrow N^*$ vertex.
The dotted curve is obtained by neglecting the off shell effects
in the $\pi N$ only calculation. Experimental data are
from Ref.~\cite{Arndt:2003fj}.}
\label{fig:gammapi0-tot}
\end{figure}

Clearly, the fit to the data needs to be improved, but is
sufficient for revealing the coupled-channels effects in a dynamical
approach. In electromagnetic pion productions, the coupled-channel effects
are in the loop integrations over the intermediate
meson-baryon states $MB$ in the following expressions
for the non-resonant amplitudes and the dressed $\gamma N \rightarrow N^*$
vertex
\begin{eqnarray}
t_{\pi N,\gamma N} = v_{\pi N,\gamma N} + \sum_{MB} t_{\pi N, MB}G_{MB}v_{MB,\gamma N}\,,
\label{eq:non-gnpn} \\
\bar{\Gamma}_{N^*,\gamma N} = {\Gamma}_{N^*,\gamma N}
+\sum_{MB} \bar{\Gamma}_{N^*,MB}G_{MB}v_{MB,\gamma N} \,.
\label{eq:nstar-gn}
\end{eqnarray}
We show the coupled-channels effects on the total cross sections
of $\gamma p \rightarrow \pi^0p, \pi^+n$ 
in Fig.~\ref{fig:gammapi0-tot}. We see that the calculated total cross 
sections (solid curves) are in good agreement with the data.
The dashed curves are obtained when the channels $ MB= \eta N$, 
$\pi\Delta$, $\rho N$, and $\sigma N$ are turned off in the 
loop integrations of Eqs.(\ref{eq:non-gnpn})-(\ref{eq:nstar-gn}).
 Clearly, the coupled-channels 
effects $\gamma N \rightarrow \eta N$, $\pi\Delta$, $\rho N$, 
$\sigma N$ $\rightarrow \pi N$ can change the cross sections by 
about 10 - 20 $\%$ in the $\Delta$ (1232) region and as much as 
50 $\%$ in the $W > $1400 MeV second resonance region.

\begin{figure}
\centering
\includegraphics[width=5cm,angle=-0]{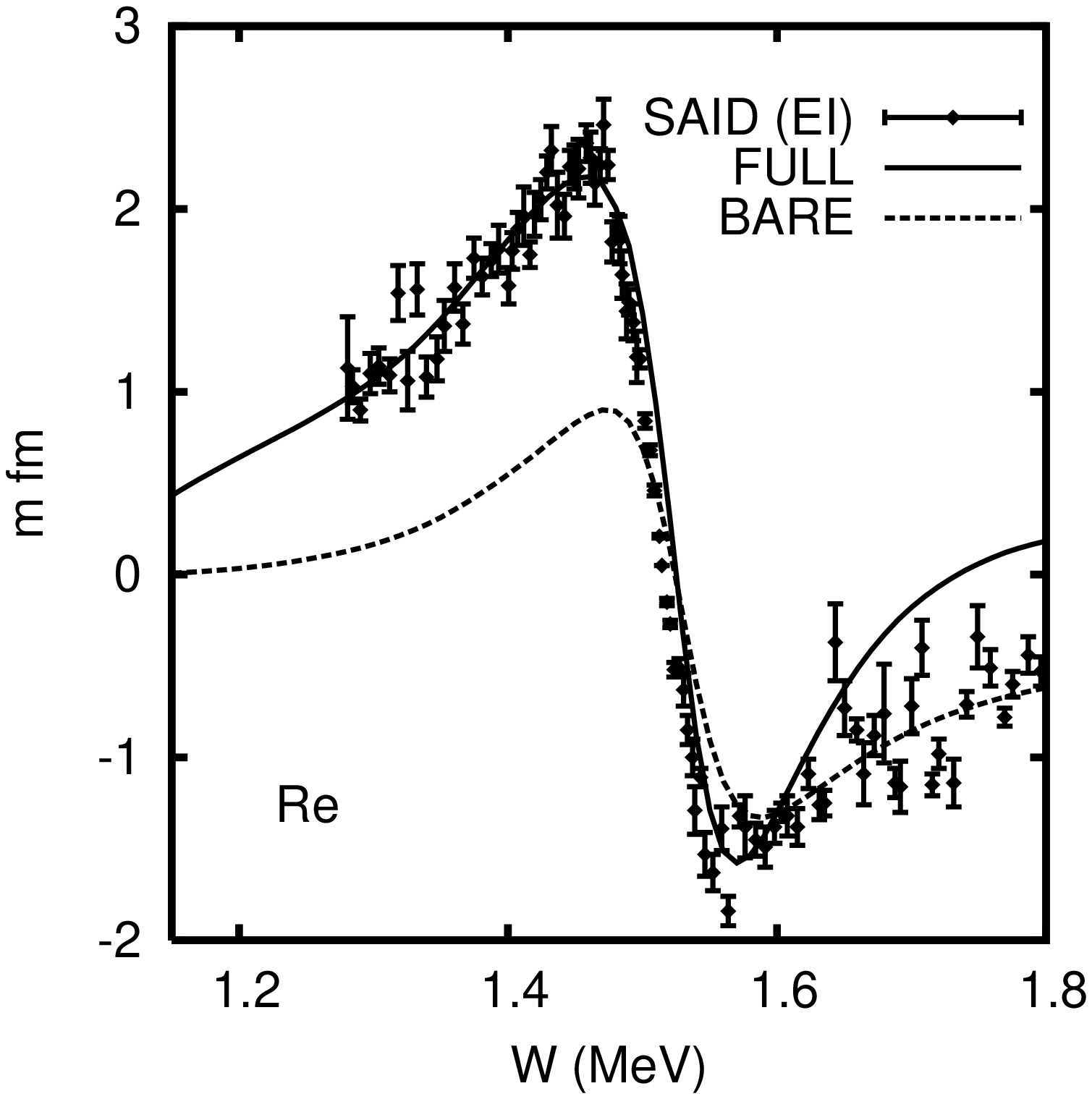}
\includegraphics[width=5cm,angle=-0]{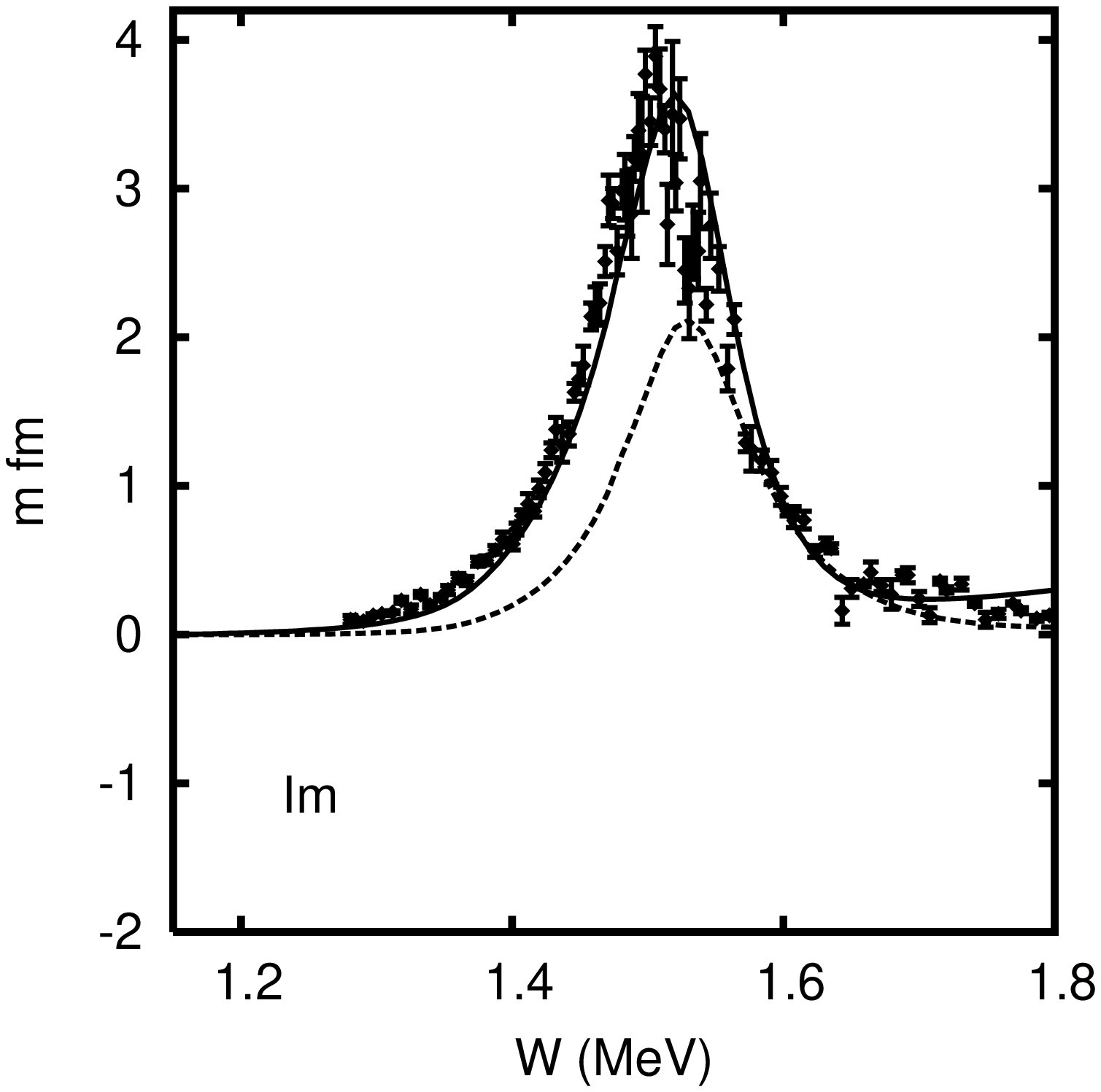}
\caption{The predicted
$\gamma N \rightarrow \pi N$ multipole amplitudes in $D_{13}$
are compared with the empirical values of SAID\cite{Arndt:2003fj}.}
\label{fig:mult-d13}
\end{figure}

The meson cloud effects, as illustrated in Fig.\ref{fig:mb-cloud},
 on several low-lying nucleon resonances 
are also investigated in Ref.\cite{JuliaDiaz:2007fa}.
In general, the resonance
parameters must be rigorously defined by the poles
 on the
unphysical sheet of complex energy plane.
This is still being pursued\cite{Suzuki:2008rp}.
 Here we only illustrate the meson cloud effect
on the $\gamma N \rightarrow \pi N$ multipoles for the $D_{13}$ partial
wave. The results are shown in
Fig.\ref{fig:mult-d13}. We see  that
the predicted
 multipole amplitudes agree well with the empirical values of 
SAID\cite{Arndt:2003fj}, and
show typical resonant shape at $W \sim 1.5 $ GeV.
Our model thus also has identified a resonance at position close to
the $N^*(1520, D_{13})$ listed by PDG. If we turn off the
meson cloud effects on the $\gamma N \rightarrow N^*$ in this partial wave,
we then get the dashed curve. Clearly, meson cloud effects are very large.

The results reviewed in this subsection are from the very first step of
performing a dynamical coupled-channel analysis of
 $\pi$ photoproduction and electroproduction
 reactions up to $W = 2 $GeV. In parallel, the investigation
of $\pi N \rightarrow \pi\pi N$ described in subsection 5.2 has also been
extended to investigate $\gamma^* N \rightarrow \pi \pi N$ reactions.
Only when the world's data of $\pi N, \gamma^* N \rightarrow \pi N, \pi\pi N$
are all included in the analysis, we can establish the $N^*$ spectrum
and their decay properties with confidence. Progress in this
is being made at EBAC.

\section{Summary and future developments}

In this article, we have reviewed the dynamical model developed  in
Refs.
\cite{Sato:1996gk,Sato:2000jf,Sato:2003rq,Matsui:2005ns,
Matsuyama:2006rp,JuliaDiaz:2006xt,JuliaDiaz:2007kz,JuliaDiaz:2007fa}
 for investigating the excitations of
$N^*$ states in $\pi N$, $\gamma N$ and $N(e,e'\pi)$ reactions.
The model Hamiltonian was constructed by using a unitary
transformation method, and had been used to construct a multi-channels and
multi-resonances reaction model. The channels considered are $\gamma N$,
$\pi N$, $\eta N$, and $\pi\pi N$ which has resonant $\pi\Delta$,
$\rho N$, and $\sigma N$ channels. The resonant amplitudes are generated
from 21 bare $N^*$ states which are renormalized by meson-baryon
scattering as required by the unitary condition. 
The model is reduced
to the well-studied Sato-Lee (SL) model when only one bare $\Delta$ state and
$\pi N$ and $\gamma N$ channels are kept in the formulation.

The detailed 
investigations\cite{Sato:1996gk,Sato:2000jf,Sato:2003rq,Matsui:2005ns}
of the $\Delta$ (1232) have  determined 
the electromagnetic $\gamma N \rightarrow \Delta$ (1232) and
the axial $A N \rightarrow \Delta$ (1232) form factors.
The meson cloud effects on these form factors are found
to be very large in the low $Q^2$ region and decreases with $Q^2$.
These form factors can be considered along with the nucleon form factors
as benchmark data for testing the predictions from hadron models
with effective degrees of freedom and LQCD.
 
The investigation of higher mass $N^*$ states is based on the
full model presented in sections 3 and 4. 
The $N^*$ parameters can be reliably determined only when
all of the available 
data of $\pi N$, $\gamma N$ and $N(e,e')$ reactions with all possible
two-particle and $\pi\pi N$ final states are fitted simultaneously.
This ambitious work, started in 2006 at EBAC, has been progressing 
well to obtain good fits to the data of $\pi N$ elastic scattering,
$\pi N \rightarrow \pi\pi N$, and $\gamma N \rightarrow \pi N$ reactions.
Important coupled-channel effects have been revealed. Large meson cloud
effects on $\gamma N\rightarrow N^*$ have also been identified. 
But more works are needed to establish the extracted $N^*$ parameters.

The current effort at EBAC is to obtain fits to the world data of
$\pi N, \gamma^* N \rightarrow \pi N, \eta N, \pi\pi N$. Staring with
the resulting $N^*$ parameters, we then focus on the $W \ge 1.7$ GeV
region by also fitting the world data of 
$\pi N, \gamma^* N \rightarrow K\Lambda, K\Sigma, \omega N$. 
The numerical strategies for handling these additional channels
have  been developed and tested. This effort is needed to face the
challenge from the complete and over complete measurements of all
independent observables of the electromagnetic production of 
$K\Lambda, K\Sigma$ reactions. These measurements are expected to be 
carried out in the next few years at JLab. Similar complete experiments
are also being developed at Mainz and Bonn.

To end of this article, we point out that the $\pi N$ data are very 
limited except the $\pi N$ elastic scattering. This could be the main source
of the uncertainties of the extracted resonance parameters.
It will be highly desirable, if more $\pi N$ reaction data can be
obtained at new hadron facility J-PARC in Japan.

\ack

We would like to thank B. Julia-Diaz, H. Kamano, A. Matsuyama, and N. Suzuki
for their collaborations on the works at EBAC. 
This work is supported by the U.S. Department of Energy, 
Office of Nuclear Physics Division, under contract No. DE-AC02-06CH11357, 
and Contract No. DE-AC05-060R23177 
under which Jefferson Science Associates operates Jefferson Lab,
and by the Japan Society for the Promotion of Science,
Grant-in-Aid for Scientific Research(c) 20540270.

\appendix

\section{Interaction Lagrangians }

The expressions of the full Lagrangians for developing
the multi-channel multi-resonance reaction model 
are given 
in Appendix A of Ref. \cite{Matsuyama:2006rp}.
In this appendix  we only give
the interaction Lagrangians for developing the SL model of
 electroweak pion production reactions. The Lagrangian 
with $\pi$, $\rho$, $\omega$, $N$, and $\Delta$ fields are\begin{eqnarray}
L_{\pi NN} &=& -\frac{f_{\pi NN}}{m_\pi}\bar{\psi}_N\gamma_\mu \gamma_5 
\vec{\tau}\psi_N\cdot \partial^\mu\vec{\phi_\pi} \,, \label{eq:L-pinn} \\
L_{\rho NN} &=& g_{\rho NN}\bar{\psi}_N[\gamma_\mu -
\frac{\kappa_\rho}{2m_N}\sigma_{\mu\nu}\partial^\nu]
\vec{\rho^\mu} \cdot \frac{\vec{\tau}}{2}\psi_N \label{eq:L-rnn} \,, \\
L_{\rho\pi\pi}&=& g_{\rho\pi\pi}[\vec{\phi_\pi}\times
\partial_\mu \vec{\phi_\pi}] \cdot \vec{\rho}^\mu\,, \\
L_{\omega NN} &=& g_{\omega NN}\bar{\psi}_N[\gamma_\mu -
\frac{\kappa_\omega}{2m_N}\sigma_{\mu\nu}\partial^\nu]
{\omega^\mu}  \psi_N\,, \\
L_{\pi N\Delta} &=& -\frac{f_{\pi N\Delta}}{m_\pi}\bar{\psi}_\Delta^\mu
\vec{T}\psi_N\cdot \partial_\mu\vec{\phi_\pi}\,. \label{eq:L-pind}
\end{eqnarray}

The effective Lagrangians for the lepton induced electroweak meson
production reaction are given as
\begin{eqnarray}
L_{eff} & =& 
\frac{4\pi \alpha}{q^2}(- \bar{e}\gamma^\mu e)j_{em,\mu}
\nonumber \\
& &
- \frac{G_F \cos\theta_c}{\sqrt{2}}[
  \bar{\nu}_e \gamma^\mu(1-\gamma_5)e j^{\dagger}_{CC,\mu}
 + \bar{e} \gamma^\mu(1-\gamma_5)\nu_e j_{CC,\mu}]
 \nonumber \\
& & - \frac{G_F}{\sqrt{2}}[
   \bar{\nu}_e\gamma^\mu(1-\gamma_5)\nu_e 
+ \bar{e}(2g_V^e\gamma^\mu - 2 g_A^e \gamma^\mu\gamma^5)e
]j_{NC,\mu}.
\end{eqnarray}
where $\alpha=1/137$, $ G_F =1.1664\times 10^{-5}$ GeV$^{-2}$
and $g_V^e = - 1/2 + 2 \sin^2\theta_W$, $g_A^e = -1/2$.
 The Weinberg angle
$\theta_W$  is 
 known empirically to be   
 $\sin^2\theta_W=0.231$ and
$\cos\theta_c = 0.974$ is the the Cabibbo-Kobayashi-Maskawa (CKM) coefficient.

The electromagnetic current ($j^\mu_{em}$), weak charge current
($j^\mu_{cc}$) and weak neutral current ($j^\mu_{nc}$)
are written with the iso-vector vector current $V^\mu_i$,
axial vector current $A^\mu_i$ and  iso-scalar vector current
$V^\mu_{is}$ as
\begin{eqnarray}
j^\mu_{em} &=& V^\mu_3+ V^\mu_{is} \,, \\
j^\mu_{cc} &=& (V^\mu_1+i V^\mu_2) - (A^\mu_1 + i A^\mu_2) \,, \\
j^\mu_{nc} &=& (1-2 \sin^2\theta_W) j^\mu_{em} - V^\mu_{is}
-A^\mu_3 \, .
\end{eqnarray}
Here we have neglected the strangeness content of the nucleon.
The iso-vector vector current$\vec{V}_\mu$ and iso-scalar vector current
are $V_\mu^{is}$
\begin{eqnarray}
\fl
\vec{V}_\mu   = 
 \bar{\psi}_N[ F_{1V}\gamma_\mu -\frac{F_{2V}}{2m_N}\sigma_{\mu\nu}
\partial^\nu ]\frac{\vec{\tau}}{2}\psi_N
+ 
\vec{\phi}_\pi\times \partial_\mu \vec{\phi}_\pi
\nonumber \\
 +
 \frac{f_{\pi NN}}{m_\pi}[
(\bar{\psi}_N\gamma_\mu \gamma_5 \vec{\tau} \psi_N)\times \vec{\phi}_\pi]
+
\frac{g_{\omega\pi\gamma}}{m_\pi}
\epsilon_{\alpha\mu\gamma\delta} \vec{\phi}_\pi
(\partial^\gamma \omega^\delta)\partial^\alpha\,,\\
\fl
{V}^{is}_\mu   = 
 \bar{\psi}_N[ F_{1S}\gamma_\mu -\frac{F_{2S}}{2m_N}\sigma_{\mu\nu}
\partial^\nu ]\frac{1}{2}\psi_N
+ 
\frac{g_{\rho\pi\gamma}}{m_\pi}
\epsilon_{\alpha\mu\gamma\delta}\vec{\phi}_\pi\cdot
(\partial^\gamma \vec{\rho^\delta})\partial^\alpha\,.
\end{eqnarray}
The axial vector current needed to construct our model is given as
\begin{eqnarray}
\fl
\vec{A}^{\mu}  =  g_A \bar{N} \gamma^\mu \gamma_5\frac{\vec{\tau}}{2} N
     - f_{\rho\pi A}\vec{\rho}^\mu \times \vec{\pi}
     - F_\pi \partial^\mu \vec{\pi}.
\end{eqnarray}
Here $F_\pi=93$ MeV is the pion decay constant, and $g_A=1.26$ is the
nucleon axial coupling constant.
The iso-vector vector  $N\Delta$ transition current
are parametrized in the following form
\begin{eqnarray}
\vec{V}_\nu  &=& -i \bar{\psi}_{\Delta}^\mu 
\Gamma^{V}_{\mu\nu} \vec{T} \psi_N  + (h.c.)\,.
\end{eqnarray}
The  matrix element of $N\Delta$ current
between an $N$ with momentum $p$ and a $\Delta$ with momentum
$p_{\Delta}$ can be written explicitly as
\begin{eqnarray}
\fl
 \Gamma^{V}_{\mu\nu}
  =   \frac{m_\Delta+m_N}{2m_N}\frac{1}{(m_\Delta+m_N)^2 - q^2}
 \nonumber \\
  \times [(G_M-G_E)3\epsilon_{\mu\nu\alpha\beta}P^\alpha q^\beta
 \nonumber \\
  + G_E i\gamma_5 \frac{12}{(m_\Delta - m_N)^2 - q^2}
  \epsilon_{\mu\lambda\alpha\beta}P^\alpha q^\beta
  \epsilon^{\lambda}_{\ \ \nu\alpha\delta}p_\Delta^\gamma q^\delta
 \nonumber \\
  + G_C i\gamma_5 \frac{6}{(m_\Delta-m_N)^2 - q^2}
    q_\mu (q^2 P_\nu - q\cdot P q_\nu)],
\end{eqnarray}

The expression for the  $N\Delta$ transition axial vector current is
given in Eqs. (\ref{eq:d-axial1})-(\ref{eq:d-axial2}).

\section{Multipole amplitudes of the pseudoscalar meson production}

Here we summarize the formula related the matrix elements
$J^\mu_\alpha$ ($\alpha=em,cc,nc$)
 in Eqs. (\ref{eq:resp-t})-(\ref{eq:resp-0}) to
the CGLN amplitudes $F_\alpha$ and and multipole amplitudes.
Recovering the spin indices of $J^\mu_\alpha$, we have
\begin{eqnarray}
\chi^{\dagger}_{s'} F_\alpha \chi_s
 & = & - \frac{m_N}{4\pi E} <\pi N(s')|J^\mu_\alpha|N(s)> \epsilon_\mu,
\end{eqnarray}
where $s,s'$ are spin quantum number of nucleon.
$F_\alpha$ is  further written as sum of the contributions of vector and
axial vector currents.
\begin{eqnarray}
F_{em}  & = & F_{em}^{V}\label{eq:fem}\,, \\
F_{cc}  & = & F_{cc}^{V} - F_{cc}^{A}\,, \label{eq:fcc} \\
F_{nc}  & = & F_{nc}^{V} - F_{nc}^{A}\,. \label{eq:fnc}
\end{eqnarray}

For electromagnetic reaction, the amplitude($F_{em}$) 
amplitudes is related to the  CGLN amplitude\cite{Chew:1957tf} as 
\begin{eqnarray}
 e F_{em} & = & F_{CGLN},
\end{eqnarray}
The amplitudes are related to those of Ref. \cite{Adler:1968tw} as
\begin{eqnarray}
  F_{\alpha} & = & \frac{M_N}{4\pi E}F_{\alpha}(Adler),
\end{eqnarray}
where $E$ is center of mass energy of pion-nucleon system.

The spin structure of the vector $F^V$ and axial vector $F^A$ amplitudes
in Eqs.(\ref{eq:fem})- (\ref{eq:fnc}) for each of
 $em, cc, nc$ currents
can be parametrized  as 
\begin{eqnarray}
\fl
F^V  =   - i\vec{\sigma}\cdot \vec{\epsilon}_\perp F_1^V
 -  \vec{\sigma}\cdot \hat{k} \vec{\sigma}\cdot\hat{q}\times\vec{\epsilon}_\perp
                                                   F_2^V
 - i\vec{\sigma}\cdot\hat{q}\hat{k}\cdot\vec{\epsilon}_\perp F_3^V
 - i\vec{\sigma}\cdot\hat{k}\hat{k}\cdot\vec{\epsilon}_\perp F_4^V
 \nonumber \\
 - i\vec{\sigma}\cdot\hat{q}\hat{q}\cdot\vec{\epsilon} F_5^V
 - i\vec{\sigma}\cdot\hat{k}\hat{q}\cdot\vec{\epsilon} F_6^V
 + i \vec{\sigma}\cdot\hat{k} \epsilon_0 F_7^V 
 + i \vec{\sigma}\cdot\hat{q}\epsilon_0 F_8^V\,,
  \label{fvec}
\end{eqnarray}
where $\vec{\epsilon}_\perp = \hat{q}\times(\vec{\epsilon} \times \hat{q})$
and
\begin{eqnarray}
\fl
F^A  =   - i\vec{\sigma}\cdot\hat{k}\vec{\sigma}\cdot \vec{\epsilon}_\perp F_1^A
 -  \vec{\sigma}\cdot\hat{q}\times\vec{\epsilon}_\perp
                                                   F_2^A
 - i\vec{\sigma}\cdot\hat{k}
    \vec{\sigma}\cdot\hat{q}\hat{k}\cdot\vec{\epsilon}_\perp F_3^A
 - i\hat{k}\cdot\vec{\epsilon}_\perp F_4^A
 \nonumber \\ 
 - i\vec{\sigma}\cdot\hat{k}
    \vec{\sigma}\cdot\hat{q}\hat{q}\cdot\vec{\epsilon} F_5^A
 - i\hat{q}\cdot\vec{\epsilon} F_6^A
 + i \epsilon_0  F_7^A
 + i \vec{\sigma}\cdot\hat{k}\vec{\sigma}\cdot\hat{q} \epsilon_0 F_8^A\,.
  \label{faxi}
\end{eqnarray}
Here $\vec{q}$ and $\vec{k}$ are momentum transfer to nucleon and
pion momentum in the center of mass system.
We defined $F^A$ simply as $\vec{\sigma}\cdot\hat{k}F^V$.

Finally the  amplitudes $F_i^V, F_i^A$ are expressed
in terms of multipole amplitudes $E_{l\pm}^{V,A},M_{l\pm}^{V,A},
S_{l\pm}^{V,A}$ and $L_{l\pm}^A$.
\begin{eqnarray}
F_1^V   =   \sum_l[
  P_{l+1}' E_{l+}^V  + P_{l-1}'     E_{l-}^V 
+ lP_{l+1}'M_{l+}^V + (l+1)P_{l-1}' M_{l-}^V]\,,\\
F_2^V   =   \sum_l[
                    (l+1)P_l'M_{l+}^V + lP_l' M_{l-}^V]\,,\\
F_3^V   =   \sum_l[
  P_{l+1}'' E_{l+}^V  + P_{l-1}''   E_{l-}^V 
- P_{l+1}'' M_{l+}^V  + P_{l-1}''   M_{l-}^V]\,,\\
F_4^V   =   \sum_l[
- P_{l}'' E_{l+}^V  - P_{l}''   E_{l-}^V 
+ P_{l}'' M_{l+}^V  - P_{l}''   M_{l-}^V]\,,\\
F_5^V   =   \sum_l[
  (l+1) P_{l+1}' L_{l+}^V  -  l  P_{l-1}' L_{l-}^V]\,, \\
F_6^V   =   \sum_l[
 -(l+1) P_{l}' L_{l+}^V  +  l  P_{l}' L_{l-}^V]\,, \\
F_7^V   =   \sum_l[
 -(l+1) P_{l}' S_{l+}^V  +  l  P_{l}' S_{l-}^V]\,, \\
F_8^V   =   \sum_l[
  (l+1) P_{l+1}' S_{l+}^V  -  l  P_{l-1}' S_{l-}^V]\,,
\end{eqnarray}
and
\begin{eqnarray}
F_1^A   =   \sum_l[
  P_{l}' E_{l+}^A  + P_{l}'     E_{l-}^A 
+ (l+2)P_{l}'M_{l+}^A + (l-1)P_{l}' M_{l-}^A]\,,\\
F_2^A   =   \sum_l[
  (l+1)P_{l+1}'M_{l+}^A + lP_{l-1}' M_{l-}^A]\,,\\
F_3^A   =   \sum_l[
  P_{l}'' E_{l+}^A  + P_{l}''   E_{l-}^A 
+ P_{l}'' M_{l+}^A  - P_{l}''   M_{l-}^A]\,,\\
F_4^A   =   \sum_l[
- P_{l+1}'' E_{l+}^A  - P_{l-1}''   E_{l-}^A 
- P_{l+1}'' M_{l+}^A  + P_{l-1}''   M_{l-}^A]\,,\\
F_5^A   =   \sum_l[
 -(l+1) P_{l}' L_{l+}^A  +  l  P_{l}' L_{l-}^A] \,,\\
F_6^A   =   \sum_l[
  (l+1) P_{l+1}' L_{l+}^A  -  l  P_{l-1}' L_{l-}^A]\,, \\
F_7^A   =   \sum_l[
  (l+1) P_{l+1}' S_{l+}^A  -  l  P_{l-1}' S_{l-}^A] \,,\\
F_8^A   =   \sum_l[
 -(l+1) P_{l}' S_{l+}^A  +  l  P_{l}' S_{l-}^A] .
\end{eqnarray}
$P_L(x)$ is Legendre function and $x=\hat{k}\cdot\hat{q}$.
In addition to the normalization of the amplitude it is noticed that
$L_{l\pm}^A,S_{l\pm}^A$ differ from those of Adler.

The multipole amplitudes are easily calculated from the
helicity-LSJ mixed representation (Eqs. (C.1) and (C.2) of Ref. \cite{Matsuyama:2006rp}).
We express
\begin{eqnarray}
<j_\pm|F_\alpha|\lambda,\lambda_N>
 & = & - \frac{m_N}{4\pi
 E}<(l1/2)j|J_\alpha\cdot\epsilon_\lambda|\lambda_N>\,,
\end{eqnarray}
where $j_\pm = j \pm 1/2$. The partial wave expansion
of the pion production current is given as
\begin{eqnarray}
\fl
<(l1/2)j|J_\alpha\cdot\epsilon_\lambda|\lambda_N>
  = 
2\pi \sum_{\lambda_N'}\int d(\cos \theta)
\sqrt{\frac{2l+1}{2j+1}}(l,0,1/2,-\lambda'_N|j,-\lambda'_N) \nonumber \\
\fl
         \times <\pi(\vec{k}),N(-\vec{k},s_N'=-\lambda'_N)|J_\alpha\cdot
	  \epsilon_\lambda| N(-\vec{q},s_N=-\lambda_N)>
 d^{(j)}_{\lambda-\lambda_N,-\lambda'_N}(\theta).
\end{eqnarray}
Here we have chosen $\vec{q}=|\vec{q}|(0,0,1)$, $\vec{k}=|\vec{k}|(\sin\theta,0,\cos\theta)$ and 
$\epsilon_{\pm 1}^\mu = (0,\mp
1/\sqrt{2},-i/\sqrt{2},0)$, $\epsilon_0^\mu=(0,0,0,1)$ and
$\epsilon_{0_t}^\mu=(1,0,0,0)$. After some derivation, we obtain the following relations:
\begin{eqnarray}
\fl
E_{l+}^V  =  \frac{1}{4\pi i (l+1)}[
                           <j_+|F^V|1,1/2>
 - \sqrt{\frac{l}{l+2}}    <j_+|F^V|1,-1/2>]\,, \\
\fl
E_{l-}^V   =  \frac{1}{4\pi i l}[ 
                         - <j_-|F^V|1,1/2>
 - \sqrt{\frac{l+1}{l-1}}  <j_-|F^V|1,-1/2>]\,, \\
\fl
M_{l+}^V   =  \frac{1}{4\pi i (l+1)}[  
                           <j_+|F^V|1,1/2>
 + \sqrt{\frac{l+2}{l}}    <j_+|F^V|1,-1/2>]\,, \\
\fl
M_{l-}^V   =  \frac{1}{4\pi i l}[ 
                           <j_-|F^V|1,1/2>
 - \sqrt{\frac{l-1}{l+1}}  <j_-|F^V|1,-1/2>]\,, \\
\fl
L_{l+}^V  =  - \frac{\sqrt{2}}{4\pi i (l+1)}  <j_+|F^V|0,-1/2>\,,\\
\fl
L_{l-}^V  =    \frac{\sqrt{2}}{4\pi i l}      <j_-|F^V|0,-1/2>\,,\\
\fl
S_{l+}^V   =   \frac{\sqrt{2}}{4\pi i (l+1)}   <j_+|F^V|0_t,-1/2>\,,\\
\fl
S_{l-}^V   =  -\frac{\sqrt{2}}{4\pi i l}      <j_-|F^V|0_t,-1/2>\,,
\end{eqnarray}
and
\begin{eqnarray}
\fl
E_{l+}^A   =  \frac{1}{4\pi i (l+1)}[ 
                            <j_+|F^A|1,1/2>
 + \sqrt{\frac{l+2}{l}}     <j_+|F^A|1,-1/2>]\,, \\
\fl
E_{l-}^A   =  \frac{1}{4\pi i l}[ 
                           -<j_-|F^A|1,1/2>
 + \sqrt{\frac{l-1}{l+1}}   <j_-|F^A|1,-1/2>]\,, \\
\fl
M_{l+}^A   =  \frac{1}{4\pi i (l+1)}[ 
                           -<j_+|F^A|1,1/2>
 + \sqrt{\frac{l}{l+2}}     <j_+|F^A|1,-1/2>]\,, \\
\fl
M_{l-}^A   =  \frac{1}{4\pi i l}[  
                          - <j_-|F^A|1,1/2>
 - \sqrt{\frac{l+1}{l-1}}  <j_-|F^A|1,-1/2>]\,, \\
\fl
L_{l+}^A   =  - \frac{\sqrt{2}}{4\pi i (l+1)}   <j_+|F^A|0,-1/2>\,,\\
\fl
L_{l-}^A   =    \frac{\sqrt{2}}{4\pi i l}       <j_-|F^A|0,-1/2>\,,\\
\fl
S_{l+}^A   =    \frac{\sqrt{2}}{4\pi i (l+1)}   <j_+|F^A|0_t,-1/2>\,,\\
\fl
S_{l-}^A   =  - \frac{\sqrt{2}}{4\pi i l}       <j_-|F^A|0_t,-1/2>\, .
\end{eqnarray}


\begin{thebibliography}{10}
\bibitem{Edmond:note}
We use edmond's convention: \\ $\langle j'm'| t_{kq}|jm\rangle =(-1)^{2k}
  \langle j'm'|jk| mq\rangle /\sqrt{2J'+1} \langle j'\parallel t_k \parallel
  j\rangle$.
\end{thebibliography}
\end{document}